\documentclass[aps,pre,twocolumn,a4paper,10pt,notitlepage,footnbib,superscriptaddress]{revtex4-1}
\usepackage[english]{babel}
\usepackage{lmodern}
\usepackage[latin9]{inputenc}
\usepackage{endnotes}
\usepackage{amssymb,amsmath,amsfonts}
\usepackage{textcomp}

\usepackage{graphicx}
\usepackage{dcolumn}
\usepackage{bm}
\usepackage{xcolor}

\begin{document}

\title{Novel non-equilibrium steady states in multiple emulsions}

\author{A. Tiribocchi}
\affiliation{Center for Life Nano Science@La Sapienza, Istituto Italiano di Tecnologia, 00161 Roma, Italy}
\affiliation{Istituto per le Applicazioni del Calcolo CNR, via dei Taurini 19, Rome, Italy}
\author{A. Montessori}
\affiliation{Istituto per le Applicazioni del Calcolo CNR, via dei Taurini 19, Rome, Italy}
\author{S. Aime}
\affiliation{School of Engineering and Applied Sciences, Harvard University, McKay 517, Cambridge, Massachusetts 02138, USA}
\author{M. Milani}
\affiliation{School of Engineering and Applied Sciences, Harvard University, McKay 517, Cambridge, Massachusetts 02138, USA}
\affiliation{Università degli Studi di Milano,  via Celoria 16, 20133, Milano, Italy}
\author{M. Lauricella}
\affiliation{Istituto per le Applicazioni del Calcolo CNR, via dei Taurini 19, Rome, Italy}
\author{S. Succi}
\affiliation{Center for Life Nano Science@La Sapienza, Istituto Italiano di Tecnologia, 00161 Roma, Italy}
\affiliation{Istituto per le Applicazioni del Calcolo CNR, via dei Taurini 19, Rome, Italy}
\affiliation{Institute for Applied Computational Science, John A. Paulson School of Engineering and Applied Sciences, Harvard University, Cambridge, Massachusetts 02138, USA}
\author{D. Weitz}
\affiliation{School of Engineering and Applied Sciences, Harvard University, McKay 517, Cambridge, Massachusetts 02138, USA}
\affiliation{Department of Physics, Harvard University, Cambridge, Massachusetts 02138, USA}

\date{\today}

\begin{abstract}
We numerically investigate the rheological response of a non-coalescing multiple emulsion under a symmetric shear flow. We find that the dynamics significantly depends on the magnitude of the shear rate and on the number of the encapsulated droplets, two key parameters whose control is fundamental to accurately select the resulting non-equilibrium steady states. The double emulsion, for instance, attains a static steady state in which the external droplet stretches under flow and achieves an elliptical shape (closely resembling the one observed in a sheared isolated fluid droplet), while the internal one remains essentially unaffected. Novel non-equilibrium steady states arise in a multiple emulsion. Under a low/moderate shear rates, for instance, the encapsulated droplets display a non-trivial planetary-like motion that considerably affects the shape of the external droplet. Some features of this dynamic behavior are partially captured by the Taylor deformation parameter and the stress tensor. Besides a theoretical interest on its own, our results can potentially stimulate further experiments, as most of the predictions could be tested in the lab by monitoring droplets shapes and position over time.

\end{abstract}

\maketitle

\section{Introduction}

A multiple emulsion is an intriguing example of soft material in which smaller drops of an immiscible fluid are dispersed within a larger one \cite{lissant1974emulsions,kahn2006,datta2014,vladi2017,weitz}. A well-known example is the double emulsion in which, for instance,  a water/oil emulsion is dispersed in a water-continuous phase \cite{xu2006,santos2016}. Higher complex systems are emulsions made of multi-distinct inner cores (such as a triple W/O/W/O emulsion) and mono or poly-disperse droplets encapsulated in a larger one \cite{utada2005monodisperse,abate2009,zarzar2015}.

Due to their unique hierarchical structure, these systems are highly desirable in a wide number of applications, including drug delivery of chemical and biological compounds \cite{cohen1991controlled,laugel2000modulated,cortesi2002production,lamprecht2004ph,kim2004comparative}, triggered reaction and mixing \cite{chen2008one,lahann2011recent,zhao2013multiphase}, cell-based therapies \cite{zhang2013novel}, waste water treatment \cite{li1972liquid}, cosmetics \cite{lissant1974emulsions,muguet2001formulation,lee2001preparation,lee2002effective}, and food science \cite{edris2001encapsulation,benichou2002double}. Unlike rigid colloids, they possess additional shape flexibility, adjustable, for instance, by carefully modulating thickness and viscosity of the shell of fluid \cite{omi2003,chu2003}. This is a crucial requirement in many applications where a precise control of rate of permeability, as well as on mechanical stability, is necessary \cite{alex1990,lamprecht2004ph,kim2004comparative}.

Although inherently out of equilibrium, these systems can be stabilized by means of suitable surfactants adsorbed onto the droplet interfaces. 
Indeed, the design of a well-defined multiple emulsion, with controlled size and number of secondary droplets, is fundamental for the correct functioning of devices in which
inner droplets' coalescence or cross-contamination of their content must be avoided \cite{kim2011one,i2011droplet}. In this context, it is crucial to investigate the
dynamic behavior of the internal droplets, since their reciprocal interaction, mediated by the surfactant and by the surrounding fluid,  may affect the rate of release
of the cargo carried within, as well as the stability of the entire emulsion \cite{saeki2010,vladi2017}.
This is a must in high internal phase multiple emulsions (of interest in food science and cosmetics),
in which fluid interfaces occupy large portions of the system and long-range effects may dramatically affect functionality and design.

Besides their technological relevance, multiple emulsions hold a great theoretical interest, due to the capability of exhibiting non-trivial interface topologies associated with a complex hydrodynamics, especially when subject to an external flow field \cite{chen1,chen2,chen3}.

\begin{figure*}
  \includegraphics[width=.6\linewidth]{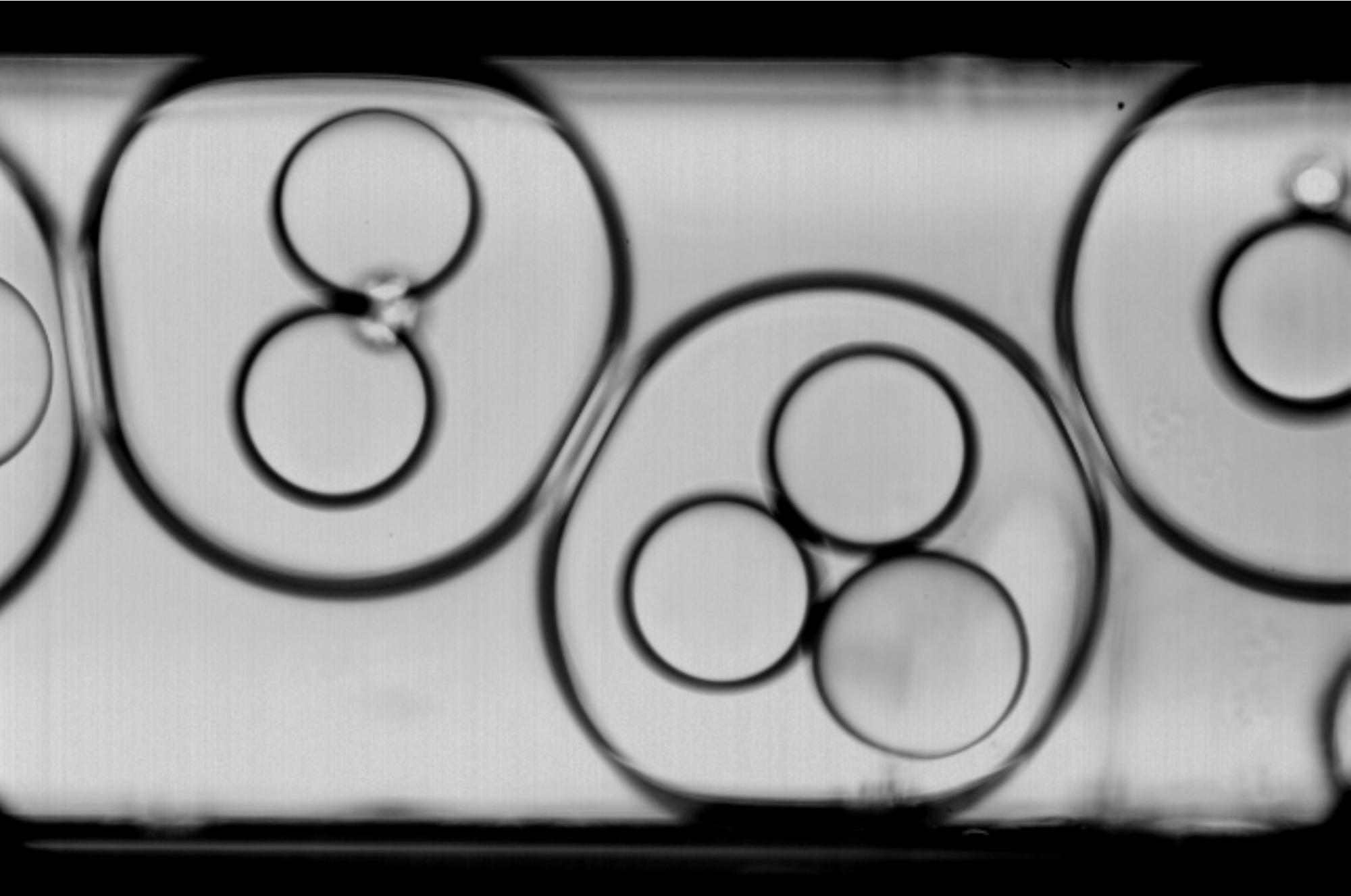}
  \caption{Double emulsion with two and three cores, fabricated in a coflowing microfluidic device \cite{utada2005monodisperse}. Here the continuous phase and the inner droplets are both water, whereas the middle phase is HFE 7500, a fluorinated oil. The interface is stabilized by adding 10\% Neat (un-dissolved) 008-FluoroSurfactant to the oil phase.}
\label{fig0} 
\end{figure*}

Yet, in spite of the impressive progress in production and design of encapsulated droplets, to date, their dynamics under an imposed flow has been only partially investigated.
While significant efforts have been addressed to understand the rheological response of single phase droplets \cite{stone,renardy,afkhami2009}, as well as of double emulsions~\cite{ha1999,wang,smith,chen1,chen2},
much less is known for higher complex systems, such as those reported in Fig.\ref{fig0}, which shows an example of a multiple emulsion with two and three cores fabricated
in a microfluidic device \cite{utada2005monodisperse}.
In the regime of low or moderate shear forces, for instance, the internal droplet of a double emulsion remains approximantely spherical and motionless at the steady state, in contrast to the external one
which attains a final ellipsoidal shape \cite{chen1,chen2} and may acquire motion. But what is the scenario if two or more inner fluid droplets are included? More specifically, 
what is their dynamics under flow? And, importantly, how do they affect shape and stability of the external droplet?

In this work, we investigate, by means of lattice Boltzmann simulations, the dynamic response of a multi-core emulsion under an externally imposed shear flow.
The basic physics of this system is captured by a multiphase field continuum model \cite{marenduzzo1,marenduzzo2}, based on a Landau free-energy description of the equilibrium properties of immiscible fluids employed to compute the thermodynamic forces (pressure tensor and chemical potential) governing the time evolution of the system \cite{degroot}. 

By varying shear rate and number of inner droplets, we observe new non-equilibrium steady states, in which the encapsulated droplets showcase a persistent periodic planetary-like motion triggered by the fluid vorticity. Such dynamics is rather robust since it occurs regardless of the initial position of the internal droplets and of their volume fraction, as long as this is sufficiently far from the close packing limit. Remarkably, this behavior leads to non-trivial modifications of the external droplet, whose steady-state shape significantly departs from the usual elliptical geometry, due to the presence of periodic local deformations occurring at its interface. These results suggest that the rheological response of a multiple emulsion is by far more complex than that of single or double emulsions, even in the regime where weak deformations are expected to occur.  

The paper is organized as follows. In the next section we describe the computational model used to simulate their rheological behavior, while in section III
we show the main numerical results. We start by investigating the rheology of a single isolated fluid droplet and afterwards we elucidate the dynamics of a double
emulsion under shear flow. Subsequently, we report the results on the non-equilibrium steady states observed in a multiple emulsion, in particular when two and three fluid droplets are encapsulated. A discussion about shape deformation and dynamic behavior of the external fluid interface is also provided.
Finally, we conclude with some remarks and perspectives.

\section{Method}

\subsection{Free energy and equations of motion}

Here we illustrate the physics and the modeling of a compound emulsion made of a suspension of immiscible fluid droplets encapsulated in larger drop.  
Such droplets are described by using a multi-phase field approach \cite{marenduzzo1,marenduzzo2,yeomans}, in which a set of scalar phase-field variables $\phi_i({\bf r},t)$, $i=1,....,N$ (where $N$ is the total number of droplets) accounts for the density of each droplet, while a vector field ${\bf v}({\bf r},t)$ describes the underlying fluid velocity.

By assuming local equilibrium \cite{degroot}, the properties of this mixture can be described by an effective coarse-grained free energy density 
\begin{equation}\label{freeE}
f= \frac{a}{4}\sum_i^N\phi_i^2(\phi_i-\phi_0)^2+\frac{k}{2}\sum_i^N(\nabla\phi_i)^2+\epsilon\sum_{i,j,i<j}\phi_i\phi_j.
\end{equation}
The first term is a double-well potential ensuring the existence of two coexisting minima, $\phi_i=\phi_0$ inside the $i$th droplet and $0$ outside.
The second term gauges the energetic cost associated to the droplet fluid interface.
The parameters $a$ and $k$ are two positive constants controlling the interfacial thickness $\xi=5\sqrt{k/2a}$ of each droplet 
and their surface tension $\sigma=\sqrt{8ak/9}$ \cite{cates1}. The last term in Eq.(\ref{freeE}) represents a soft-core repulsion whose strength
is measured by the positive constant $\epsilon$.

The dynamics of the order parameters $\phi_i({\bf r},t)$ is governed by a set of convection-diffusion equations
\begin{equation}\label{CH_eqn}
D_t\phi_i=-\nabla\cdot{\bf J}_i
\end{equation}
where $D_t=\partial/\partial_t+{\bf v}\cdot\nabla$ is the material derivative and
\begin{equation}
{\bf J}_i=-M\nabla\mu_i
\end{equation}
is the current, proportional to the product of the mobility $M$ and the gradient of the chemical potential
\begin{equation}
\mu_i\equiv\frac{\delta {\cal F}}{\delta\phi_i}=\frac{\partial f}{\partial\phi_i}-\frac{\partial_{\alpha}f}{\partial(\partial_{\alpha}\phi_i)}
\end{equation}
of the $i$th drop. Finally, ${\cal F}=\int_VfdV$ is the total free energy.

The fluid velocity ${\bf v}({\bf r},t)$ obeys the continuity and the Navier-Stokes equations which, in the incompressible limit, are
\begin{equation}\label{CNT_eqn}
\nabla\cdot{\bf v}=0,
\end{equation}
\begin{equation}\label{NAV_eqn}
  \rho\left(\frac{\partial}{\partial t}+{\bf v}\cdot\nabla\right){\bf v}=\nabla\cdot{\Pi}.
\end{equation}
In Eq.(\ref{NAV_eqn}) $\rho$ is the fluid density and $\Pi$ is the total stress tensor given by the sum of three further terms. The first one is the isotropic pressure $\Pi^{is}=-p\delta_{\alpha\beta}$ and
the second one is viscous stress $\Pi^{visc}=\eta(\partial_{\alpha}v_{\beta}+\partial_{\beta}v_{\alpha})$, where $\eta$ is the shear viscosity (Greek indexes denote Cartesian components).
Finally, the last term takes into account interfacial contributions between different phases and is given by
\begin{equation}
\Pi^{inter}=\left(f-\sum_i\phi_i\frac{\delta {\cal F}}{\delta\phi_i}\right)\delta_{\alpha\beta}-\sum_i\frac{\partial {\cal F}}{\partial (\partial_{\beta}\phi_i)}\partial_{\alpha}\phi_i.
\end{equation}
Note, in particular, that $\nabla\cdot\Pi^{inter}=-\sum_i\phi_i\nabla\mu_i$, representing the driving force due to the presence of spatially varying contributions of the
order parameters.

\subsection{Simulation details and numerical mapping}

Eqs.~(\ref{CH_eqn}), (\ref{CNT_eqn}) and (\ref{NAV_eqn}) are solved by using a hybrid numerical approach, in which the convection-diffusion equations are integrated by using
a finite difference scheme while the continuity and the Navier-Stokes equations via a lattice Boltzmann algorithm \cite{succi1,succi2,kruger,yeomans2,succi3,montessori,karlin,ansumali}.
This method has been successfully adopted to simulate a wide variety of soft matter systems, ranging from binary fluids \cite{tiribocchi,tiribocchi2} and liquid crystals
\cite{marenduzzo3,marenduzzo4,tiribocchi3,foffano,tiribocchi5} to active gels \cite{cates2,tiribocchi4}, and has been recently extended to describe the physics of
non-coalescing droplet suspensions \cite{marenduzzo1,marenduzzo2}.

\begin{figure*}
\includegraphics[width=1.\linewidth]{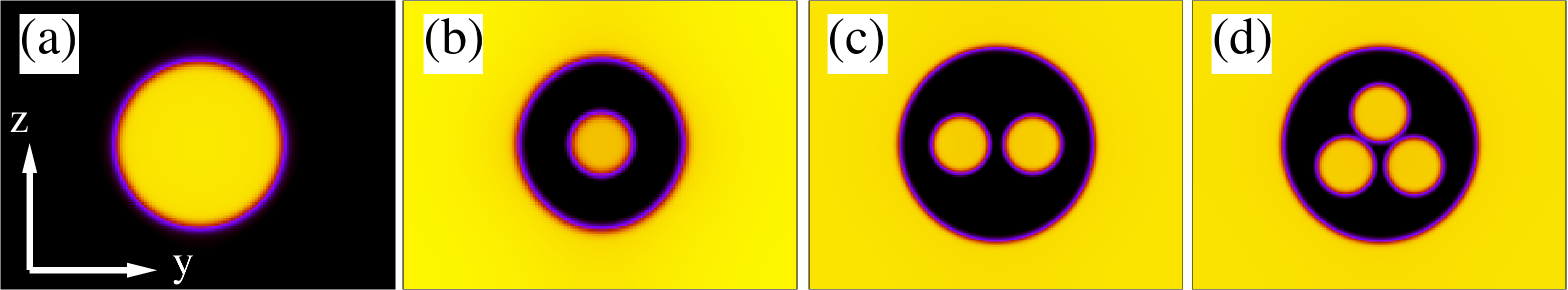}
\caption{Equilibrium profiles of different emulsions. (a) Isolated droplet, (b) double emulsion, (c) two-cores multiple emulsion, (d) three-cores multiple emulsion. Lattice dimensions are (a)-(b) $L_y=150$, $L_z=110$, (c)-(d) $L_y=220$, $L_z=170$.  Droplet radii are (a) $R=30$, (b) $R_{in}=10$, $R_{out}=30$, (c) $R_{in}=15$, $R_{out}=56$, (d) $R_{in}=15$, $R_{out}=56$. Colors correspond to the values of the order parameter $\phi$, ranging from $0$ (black) to $\simeq 2$ (yellow).}
\label{fig1} 
\end{figure*}

All simulations are performed on two dimensional rectangular lattices (see Fig.\ref{fig1}), in order to minimize interference effects due to the periodic image of the droplets.
These systems are sandwitched between two parallel flat walls, where we set no-slip conditions for the velocity field ${\bf v}$
and neutral wetting for the fields $\phi_i$. The former means that  $v_z(z=0,z=L_z)=0$, and the latter that ${\bf n}\cdot\nabla\mu_i|_{z=0,L_z}=0$ (no flux through the boundaries) and
${\bf n}\cdot\nabla(\nabla^2\phi_i)|_{z=0,L_z}=0$ (droplet interface perpendicular at the boundaries), where ${\bf n}$ is an inward normal unit vector at the boundaries.

In Fig.\ref{fig1}a an isolated isotropic fluid droplet (yellow) is initially placed at the centre of the lattice and is surrounded by a second isotropic fluid (black).
In this configuration, only one order parameter field $\phi$ is considered (i.e. $N=1$). A double emulsion (Fig.\ref{fig1}b) is produced by means of two fields $\phi_i$ ($N=2$).
One is positive (equal to $\simeq 2$) within the smaller droplet (placed at the centre of the lattice) and zero everywhere else, while the other one is positive outside the larger droplet
and zero elsewhere. Analogous setups have been employed for the other multiple emulsion, when two ($N=3$) and three ($N=4$) droplets are included.
The radii of the droplets have been chosen as follows: (a) $R=30$, (b) $R_{in}=10$ and $R_{out}=30$, (c)-(d) $R_{in}=15$ and $R_{out}=56$. The corresponding emulsion volume fraction $V_f=\frac{N\pi R_{in}^2}{\pi R_{out}^2}$ is (a) $V_f=0$, (b) $V_f\simeq 0.11$, (c) $V_f\simeq 0.15$ and (d) $V_f\simeq 0.22$.  

Starting from these initial conditions, the mixtures are first let to relax for $\simeq 5\times 10^5$ timesteps to achieve a (near) equilibrium state. Afterwards, a symmetric shear is applied, by
moving the top wall along the positive $y$-axis with velocity $v_w$ and the bottom wall along the opposite direction with velocity $-v_w$. This sets a shear rate $\dot{\gamma}=2v_w/L_z$. 
In our simulations $v_w$ ranges between $0.01$ (low shear) to $0.05$ (moderate/high shear), which means that $\dot{\gamma}$ varies between $\simeq 2\times 10^{-4}$ to $\simeq 10^{-3}$ when $L_z=110$,
and between $\simeq 1.1\times 10^{-4}$ and $\simeq 6\times 10^{-4} $ when $L_z=170$.
As in previous works \cite{chen2}, we define a dimensionless time $t^*=\dot{\gamma}(t-t_{eq})$, where $t_{eq}$ is the relaxation time after which the shear is switched on. 
Unless otherwise explicitly stated, the following thermodynamic parameters have been used: $a=0.07$, $M=0.1$, $\eta=1.67$, $k=0.1$ and $\epsilon=0.05$.
Also, throughout our simulations, time-step and lattice spacing are fixed to unit value, $\Delta x=1$, $\Delta t=1$.

By following previous studies \cite{marenduzzo1,marenduzzo2}, an approximate mapping between simulations units and physical ones
can be obtained by assuming a droplet of diameter roughly equal to $10^2$ $\mu$m immersed in a background fluid of viscosity $\simeq 10^{-2}$ $Pa\cdot s$ (assumed, for simplicity, equal to the viscosity
of the fluid inside the droplet)  and in which the surface tension $\sigma$,
equal to $\simeq 0.08$ (for $k=0.1)$ in simulations, corresponds to $\sim 0.5-1$ m$N/m$. With these parameters, a speed of $10^{-3}$ in simulation units corresponds to approximately $1$ $mm/s$ in real values. Further details are reported in the Appendix.
A dimensionless quantity capturing droplet deformation is the capillary number $Ca=\frac{v\eta}{\sigma}$, measuring the strength of the viscous forces relative
to the surface tension. If, for example, $v=0.01$, $Ca\sim 0.2$ (with $k=0.1$). In addition, the Reynolds number $Re=\rho v_{max}L/\eta$ (where $L$
is the system size) may vary from $\sim 1$ to $\sim 10$, the latter describing a regime for which inertial forces are much higher than viscous ones and the condition of laminar flow is generally not fulfilled.

\section{Results}

Here we discuss the rheological response of the emulsions shown in Fig.\ref{fig1} subject to a symmetric shear flow. To validate our model, we initially investigate the dynamics of an isolated fluid droplet and afterwards we move on to study the dynamical response of the other compound emulsions.

\subsection{Isolated fluid droplet}
\begin{figure*}
\includegraphics[width=1.\linewidth]{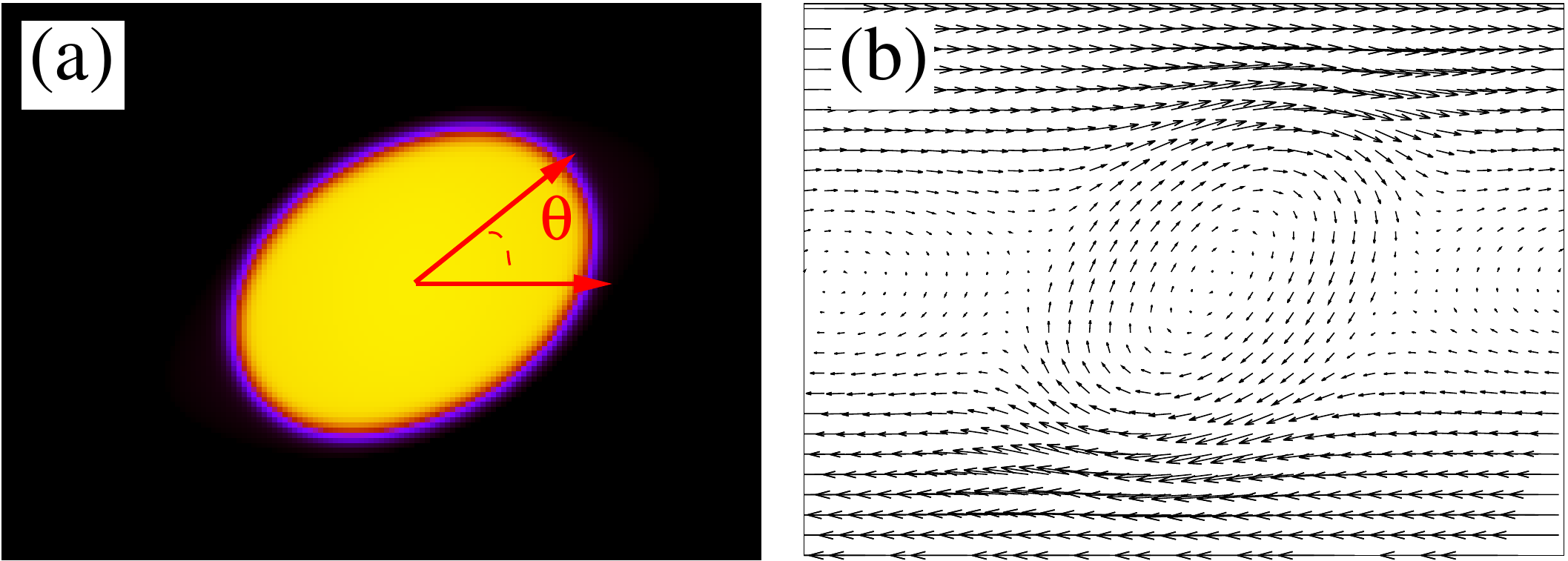}
\caption{(a) Steady state profile of an isolated droplet immersed in a second immiscible fluid subject to a symmetric shear flow  with $\dot{\gamma}\simeq 1.8\times 10^{-4}$. Also $Re\simeq 2$, $Ca\simeq 0.21$ and $k=0.1$. The angle $\theta$ indicates the direction of the droplet major axis with the shear flow. The color map is the same as that of Fig.\ref{fig1}. (b) Steady state velocity profile under shear. Intense opposite unidirectional flows are produced near the walls, whereas a much weaker fluid recirculation is observed within the droplet.}
\label{fig2} 
\end{figure*}

\begin{figure*}
\includegraphics[width=0.75\linewidth]{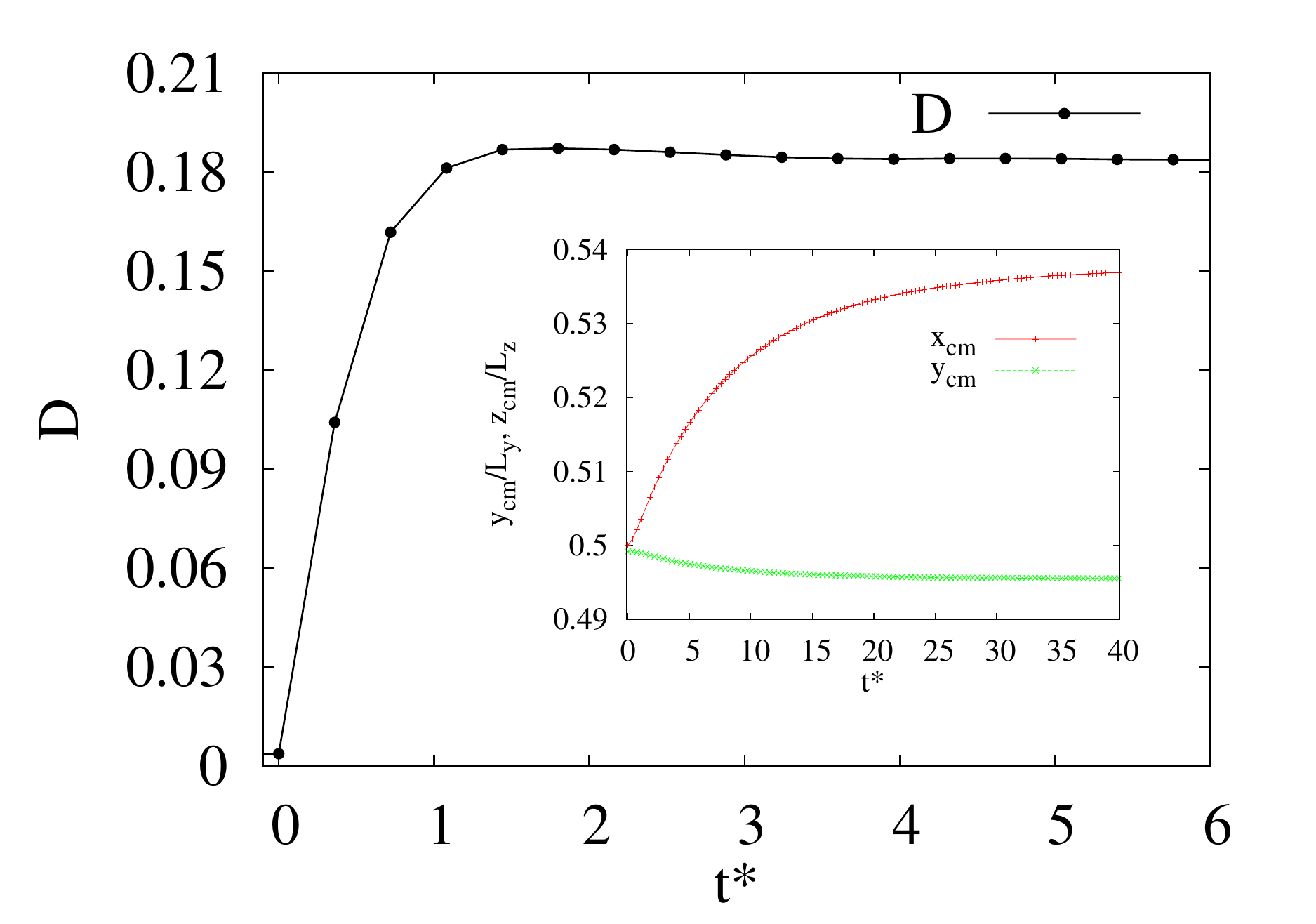}
\caption{Time evolution of the Taylor parameter $D$. Inset: Time evolution of the $y$ and $z$ components of the droplet center of mass. They are defined as $y_{cm}(t)=\frac{\sum_{y} y(t)\phi(y,z,t)}{\sum_{y} \phi(y,z,t)}$ and $z_{cm}(t)=\frac{\sum_{z}z(t)\phi(y,z,t)}{\sum_{z}\phi(y,z,t)}$, where $y=1,...,L_y$, $z=1,...,L_z$ and $\phi(y,z,t)\geq 0.1$.}
\label{fig3} 
\end{figure*}

As first benchmark test, we simulate the effect produced by a symmetric shear flow to an isolated fluid droplet surrounded by a second immiscible fluid. A well-known result is that, for low/moderate values of $\dot{\gamma}$, at the steady state the droplet attains an elliptical-like shape and alings along the imposed shear flow. 
In Fig.~\ref{fig2}a-b we show the steady state of the droplet and the corresponding fluid flow profile after imposing a shear rate $\dot{\gamma}\simeq 1.8\times 10^{-4}$ (see also movie M1 (Multimedia view)).
As expected, the droplet elongates and the major axis tilts and forms an angle of $\theta\simeq 30$ degrees with the shear direction. This is in very good agreement
with values reported in literature for $Re\simeq 2$ (see, for example,  Ref.~\cite{renardy}).
The velocity field exhibits the typical structure observed for such system, i.e. it is large and unidirectional near both walls and weaker in the centre of the lattice,
where a clockwise recirculation emerges within the droplet. The droplet position is mildly affected by the shear flow (see Fig.\ref{fig3}), which only slightly pushes the droplet rightwards with respect to the initial location. 

As long as the droplet shape remains rather well-defined (like an ellipse), one can quantify its deformation in terms of the Taylor parameter $D=\frac{a-b}{a+b}$, where
$a$ and $b$ represent the length of the major and the minor axis, respectively. It ranges between $0$ (no deformation) and $1$ (``needle'' shape). In Fig.\ref{fig3} it is shown that
$D$ attains a steady state value of $\sim 0.18$, in line with experimental values observed when $Ca\sim 0.2$ \cite{bentley,stone}. 

For higher values of $\dot{\gamma}$ (but low enough to avoid the droplet breakup \cite{zaleski}), the droplet, once more, aligns with the flow direction and attains the elliptical shape 
but with a higher deformation at the steady state. If, for example, $\dot{\gamma}\simeq 10^{-3}$ one gets $\theta\simeq 40$ degrees and $D\simeq 0.65$, values close to experimental ones for $Re\simeq 10$ \cite{renardy} (see movie M2 (Multimedia view)).

These preliminary numerical tests reproduce with very good accuracy some aspects of the dynamic response under shear of an isolated fluid droplet. In the next section we extend this study to a
Newtonian double emulsion, in which a second droplet is included within a larger one.

\subsection{Double emulsion}

\begin{figure*}
\includegraphics[width=1.\linewidth]{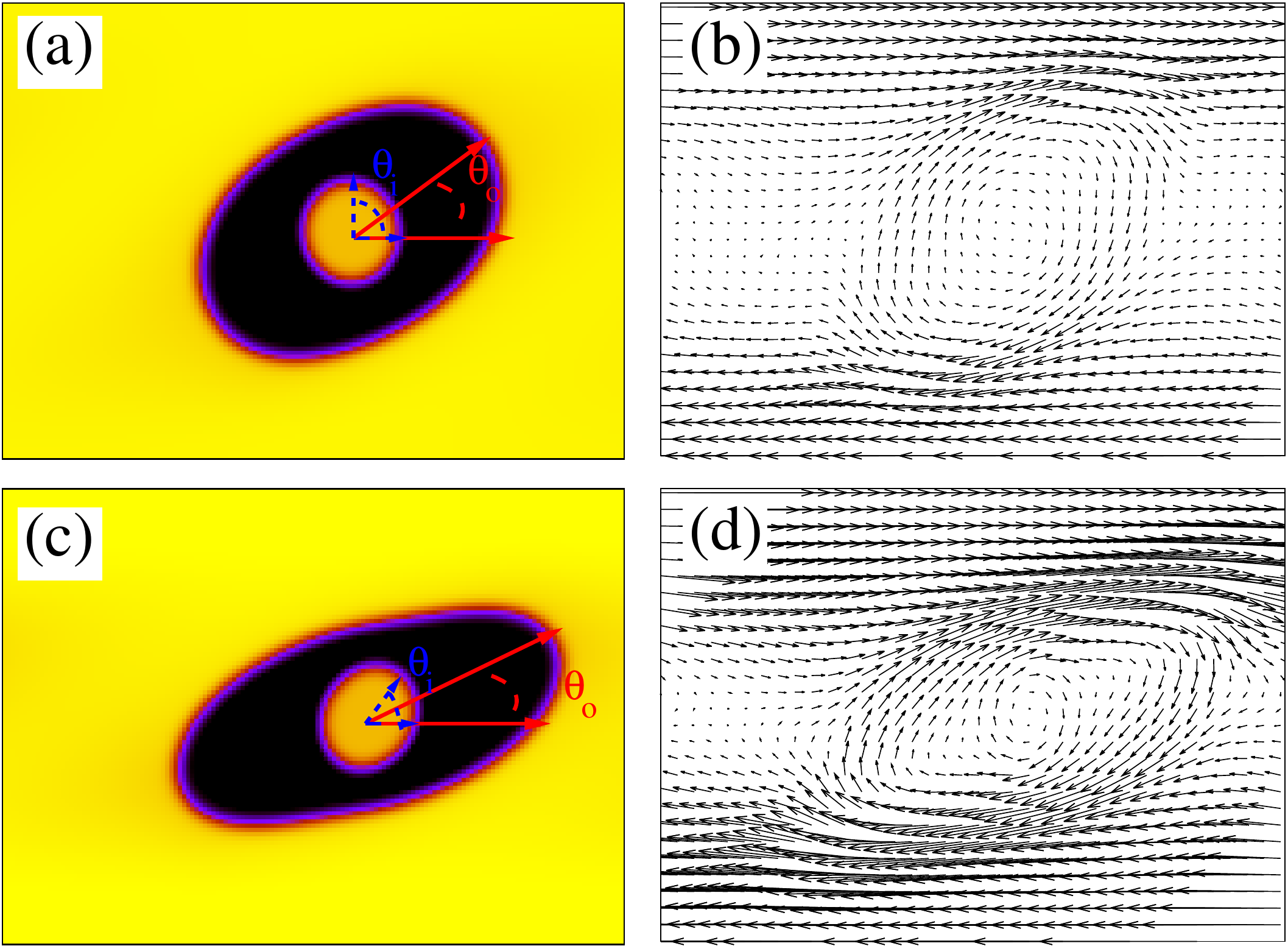}
\caption{(a)-(c) Steady state profiles of a double emulsion under a symmetric shear flow with (a) $\dot{\gamma}\simeq 1.8\times 10^{-4}$ and (c) $\dot{\gamma}\simeq 4.5 \times 10^{-4}$. Here $Re\simeq 2$, $Ca\simeq 0.21$ in (a) and  $Re\simeq 4.5$, $Ca\simeq 0.52$ in (c). (b)-(d) Steady state velocity profiles under shear.}
\label{fig4} 
\end{figure*}

\begin{figure*}
\includegraphics[width=1.\linewidth]{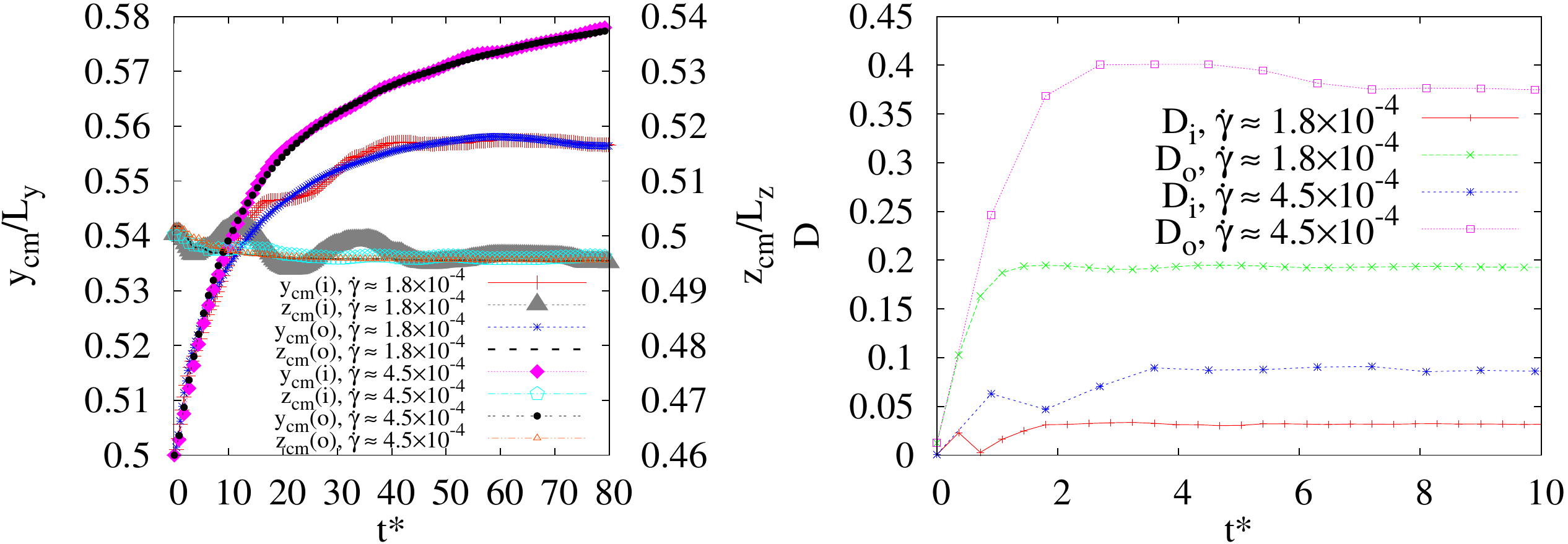}
\caption{Left: Time evolution of the $y$ and $z$ components of the droplets centers for mass in the double emulsion for two different values of $\dot{\gamma}$. In both cases, the droplet position is mildy affected by the fluid flow. Right: Time evolution of deformation parameter $D$ of the inner and the outer droplets.}
\label{fig5} 
\end{figure*}

Due to presence of an inner droplet, more complex hydrodynamics and interfacial deformations are expected with respect to the single-phase case. The effect of a moderate shear flow ($\dot\gamma\simeq 1.8\times 10^{-4}$) on a double emulsion is shown in Fig.\ref{fig4}a-b (and movie M3 (Multimedia view)). After the shear is imposed, the outer droplet is, once more, slightly advected rightwards (see Fig.\ref{fig5}, left) and, simultaneously, tilted and stretched along the shear direction, until the elliptical steady-state shape is attained at approximately $t^*\simeq 100$.

Here we measure $\theta_o\simeq 42$ degrees and $D_o\simeq 0.2$ (Fig.\ref{fig5} right, green plot), values comparable with those of the single phase droplet.
Hence, as long as $\dot{\gamma}$ is sufficiently small, the presence on the inner fluid droplet has a mild effect on the outer one, whose interface acts as an effective
``shield'' preventing deformations of the former.
Indeed, the shape of the external droplet remains almost unaltered throughout the process (we measure $\theta_i\simeq 85$ degrees and $D_i\simeq 0.03$, see Fig.~\ref{fig5}, right).
This is mainly due to the large interfacial tension (higher than that of the outer droplet)
induced by the small curvature radius, thus preventing deformations that would be favoured by the shear flow.  
A further source of shape stabilization stems from the (weak) vorticity formed within the smaller droplet (in addition to the larger one mainly located in the layer between the droplets,
see Fig.\ref{fig4}b), an effect known to inhibit deformations produced by the shear stress \cite{rallison,hakimi}.

Doubling the shear rate can produce substantial shape deformations of the inner droplet as well as of the outer one (movie M4 (Multimedia view)). In Fig.\ref{fig4}c-d we show the steady state of a double emulsion when
$\dot{\gamma}\simeq 4.5\times 10^{-4}$. Here we got $\theta_o\simeq 30^o$ and $\theta_i\simeq 40^o$, while $D_i\simeq 0.1$ and $D_o\simeq 0.4$ (Fig.~\ref{fig5} right). Once again, due to its higher surface tension,
the inner droplet is less deformed than the outer one but, unlike the the previous case, a visible rounded clockwise recirculation, clearly distinct from the large elliptical one, forms
inside.

Note that increasing $\dot{\gamma}$ produces a temporary peak in $D$, soon after the shear force is switched on. While, for low values of shear rate, droplet elongation and alignment to the flow
direction occurs gradually, for high values it stretches rather abruptly and later on relaxes towards its steady state shape. As also observed in previous works \cite{chen1,chen2}, this
initial deformation overshoot is necessary to overcome the additional inertia displayed by the droplet after an intense stretching.

Despite its comparatively simple design, the double emulsion displays a non-trivial rheological behavior, in which interface deformations and shape changes
crucially depend on the elasticity and on the complex structure of the fluid velocity.

A largely unexplored physics is that of higher complex multiple emulsions, in which, for instance, two (or more) smaller fluid droplets are included within a larger external one.  The next chapter is precisely dedicated to investigate the rheology of such system in the presence of low/moderate shear flows.

\subsection{Higher complex states: Multiple emulsion}

\subsubsection{Non-equilibrium steady states}

We first consider likely the simplest example of a multiple emulsion, namely two collinear fluid droplets located symmetrically with respect to an axis, parallel to $z$, passing through the center of mass of a surrounding larger droplet (see Fig.~\ref{fig1}c). Despite its essential design, a non-trivial rheological behavior emerges when subject to a shear flow.

\begin{figure*}
\includegraphics[width=1.\linewidth]{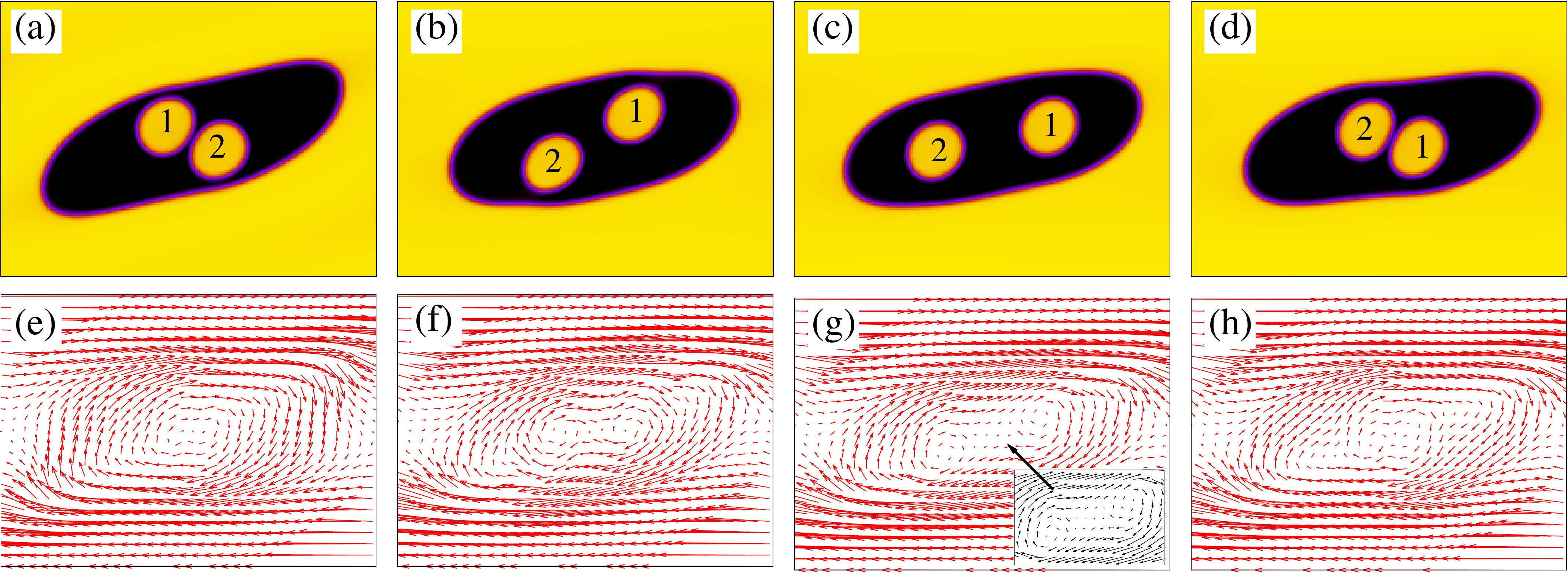}
\caption{Top row: Steady state profiles of the fields $\phi_i$ of a two-droplet oscillatory dynamics under a symmetric shear flow with $\dot{\gamma}\simeq 3\times 10^{-4}$. Snapshots are taken at (a) $t^*=3$,  (b) $t^*=7.8$, (c) $t^*=12$, (d) $t^*=18$. Here $Re\simeq 2.6$, $Ca\simeq 0.21$. Bottom: Steady state velocity profiles under shear. In (g) a zoom of the two weak recirculations rotating clockwise formed nearby the inner droplets.}
\label{fig6} 
\end{figure*}

\begin{figure*}
\includegraphics[width=1.\linewidth]{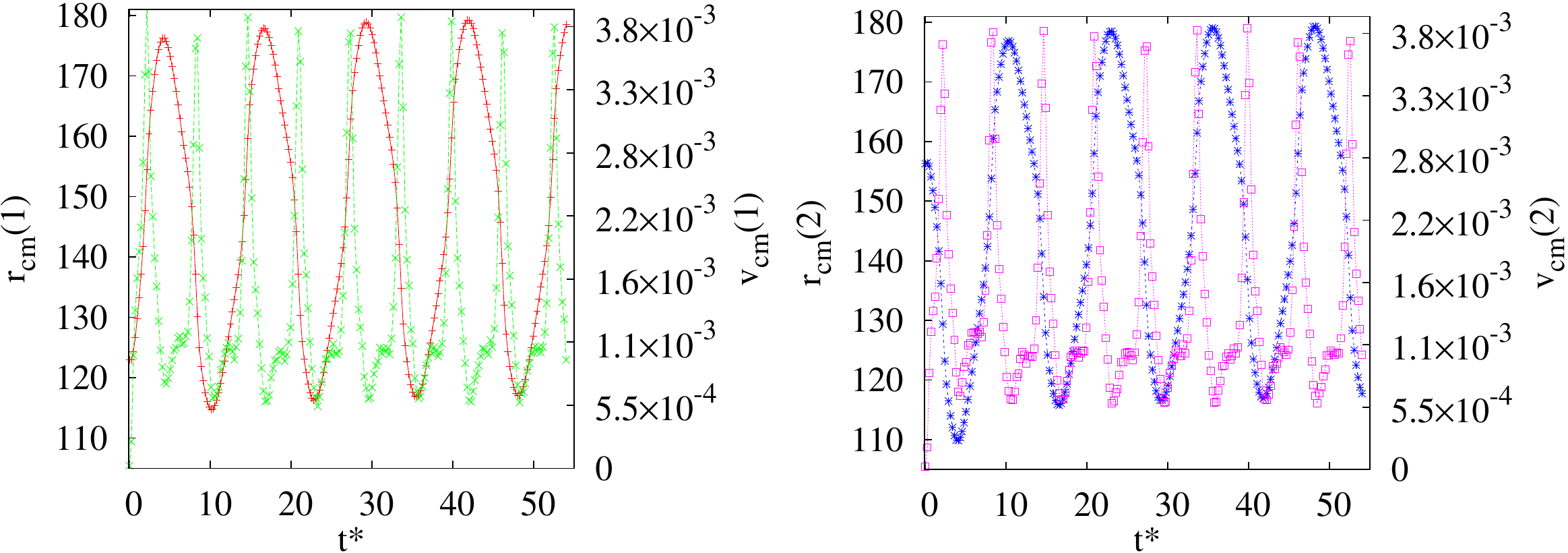}
\caption{Time evolution of the center of mass $r_{cm}$ and of its speed $v_{cm}$ when $\dot{\gamma}\simeq 3\times 10^{-4}$ for the droplet 1 (left) and droplet 2 (right) of Fig.~\ref{fig6}. Colors are as follows:  Red (plusses) and green (crosses) correspond, respectively, to center of mass and speed of droplet 1, while blue (asterisks) and magenta (squares) correspond, respectively, to center of mass and speed of droplet 2.}
\label{fig7} 
\end{figure*}

Once a moderate shear is switched on, the two inner cores acquire motion, initially proceeding along opposite directions (Fig.~\ref{fig6}a-b) and, later on, rotating periodically clockwise around the center of mass of the outer droplet by following roughly elliptical orbits (Fig.~\ref{fig6}c-d, and movie M5 (Multimedia view)). As in a typical periodic motion, internal droplets attain the minimum speed (local minima of green and magenta plots of Fig.~\ref{fig7}) at the points of inversion of motion, while the highest speed is achieved halfway (maxima and minima of red and blue plots of Fig.~\ref{fig7}). 

Such planet-like oscillatory motion, observed during the transient dynamics and persistently at the steady state, is primarily caused by the confined geometry in which the internal droplets are constrained move and, likewise, by the purely droplet-droplet repulsive interaction (an effect captured by the term proportional to $\epsilon$ in the free energy) combined with a non-trivial structure of the internal velocity field. Indeed, unlike the double emulsion, it exhibits a large fluid recirculation near the interface of the extenal droplet and two temporary recirculations within the emulsion, appearing faraway when the droplets invert their motion and merging into a single one when they are sufficiently close to each other.

Intriguingly, this dynamics produces significant effects on the external droplet shape. Although, like in the double emulsion, the droplet elongates and aligns along the direction imposed by the shear, at the steady state it shows periodic shape deformations characterized by local interfacial bumps, more intense when internal droplets approach the interface of the external one. When this occurs, the shape of the external droplet considerably departs from the typical ellipsoidal one observed in single and double emulsions, a geometry only temporary restored when internald droplets are far from the external interface.

These results suggest a scenario in which i) novel non-equilibrium steady states emerge whenever a multiple emulsion is subject to a shear flow, and ii) non-trivial deformations of the external interface emerge as a result of the internal droplets motion. But how robust is this dynamics?  And, how does it depend on the arrangement of the internal droplets? Is the parameter $D$ still a reliable quantity to capture droplet shape deformations even for moderate values of shear rates?

To partially address these questions, we study the dynamics under a moderate shear flow of a multiple emulsion in which three fluid droplets are encapsulated (see Fig.\ref{fig1}d for the initial condition). The effect of the shear is overall similar to the previous case. During the transient dynamics, the external droplet elongates and stretches along the shear flow, while the internal ones acquire a clockwise rotating motion around the center of mass of the former. 

\begin{figure*}
\includegraphics[width=1.\linewidth]{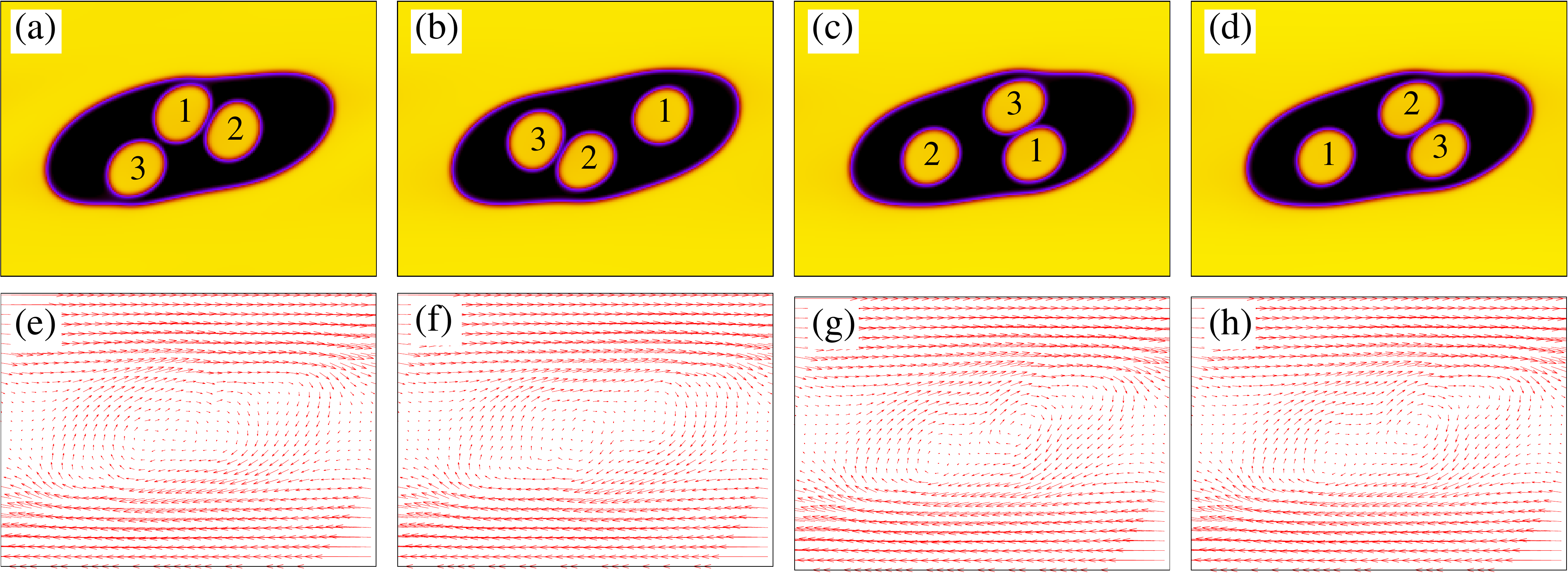}
\caption{Top row: Steady state profiles of the fields $\phi_i$ of a three-droplet oscillatory dynamics emulsion under a symmetric shear flow with $\dot{\gamma}\simeq 3\times 10^{-4}$. Snapshots are taken at (a) $t^*=6$, (b) $t^*=12$, (c) $t^*=18$, (d) $t^*=27$. Here $Re\simeq 6.6$, $Ca\simeq 0.52$. Bottom: Steady state velocity profiles under shear.}
\label{fig9} 
\end{figure*}

At the steady state, such motion becomes, once again, periodic (see Fig.~\ref{fig9}), with the internal droplets persistently moving along elliptical trajectories dragged by the large fluid recirculation formed near the external interface, where local deformations occur more frequently (see movie M6 (Multimedia view)). Further weak fluid vortices also appear mainly located around the interface of the internal droplets. Hence, as long as $\dot{\gamma}$ is low enough to prevent the emulsion rupture, the periodic motion of the internal droplets is preserved, regardless of the droplets arrangememt. Even so, it is worth noting that increasing the number of internal droplets causes a sensible decrease of the tilt angle of the external droplet at the steady state. Indeed, we measure $\theta_o\simeq 27^o$ for the two-cores emulsion and $\theta_o\simeq18^o$ for the three-cores one. 

\subsubsection{Interface deformation and stress tensor}

\begin{figure*}
\includegraphics[width=0.75\linewidth]{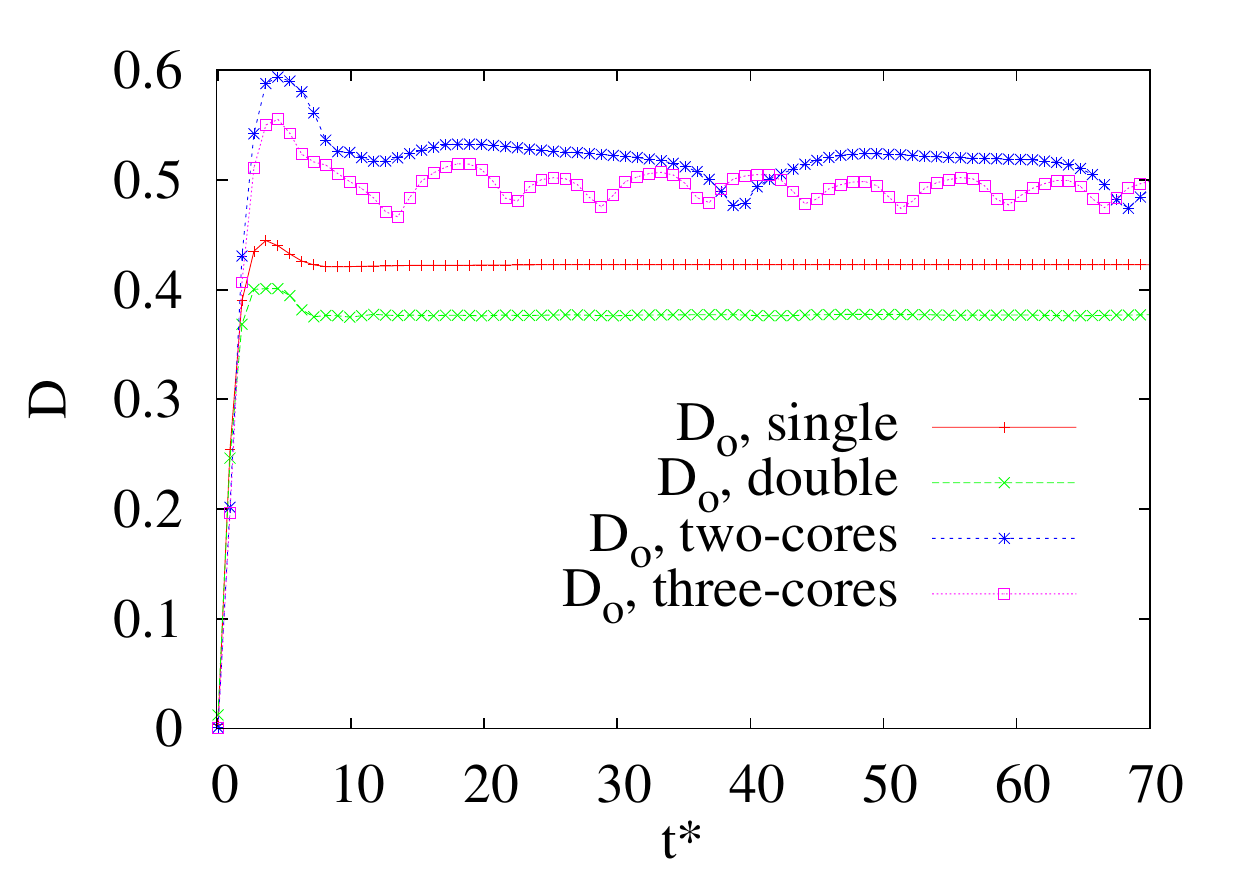}
\caption{Time evolution of the deformation $D$ of the external droplet when $\dot{\gamma}\simeq 4.5\times 10^{-4}$. Periodic oscillations emerge when two and three droplets are included.}
\label{fig8} 
\end{figure*}

An estimate of the shape deformation of the extenal droplet can be obtained by looking at the time evolution of the parameter $D$. Interestingly, while for a single and a double emulsions $D$ rapidly attains a constant value, in a multiple emulsion the periodic motion of the internal droplets leaves a tangible signature on it, which now displays periodic oscillations, more frequent as the number of the encapsulated droplets augments (see Fig.~\ref{fig8}). However, such behaviour provides only a partial knowledge about the correct shape of the external droplet, since it only captures a periodic elongation/contraction mechanism but misses the local interfacial bumps.

\begin{figure*}
\includegraphics[width=1.\linewidth]{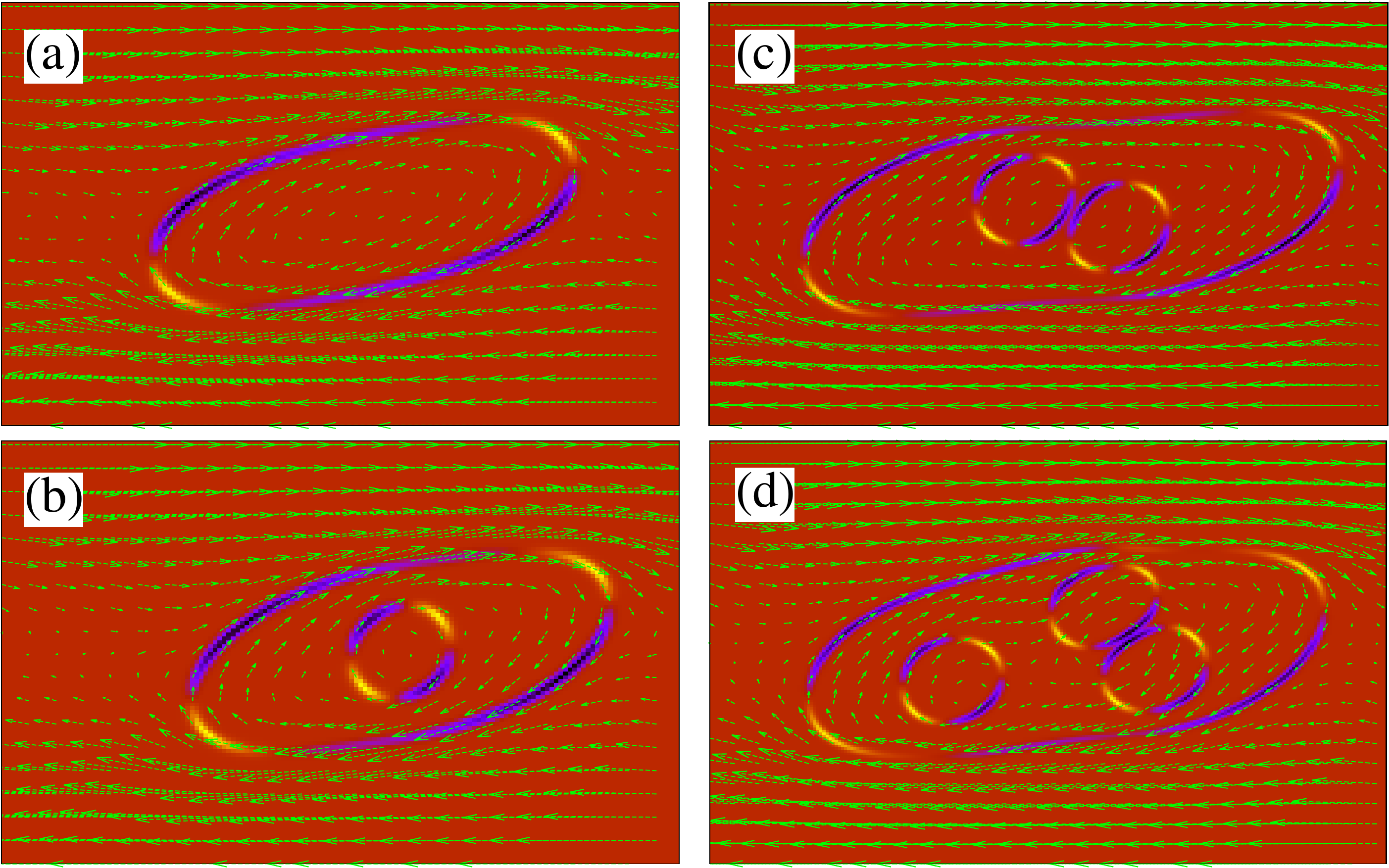}
\caption{Contour plot of $\Pi_{yz}$ at the steady state for (a) isolated droplet, (b) double emulsion, (c) two-droplet monodisperse emulsion and (d) three-droplet monodisperse emulsion. Color map ranges from $2\times 10^{-2}$ (yellow) to $-2\times 10^{-2}$ (black). Negative values at the interface depend on the sign of derivative term in the stress tensor, proportional to $\partial_y\phi\partial_z\phi$.  Arrows represent the velocity field of the fluid.}
\label{fig12} 
\end{figure*}

Further insights can be gained by computing the in plane component of the stress tensor $\Pi_{yz}$. Major contributions stem from the off diagonal term of $\Pi_{yz}^{inter}$ and from the cross derivative
terms of $\Pi_{yz}^{visc}$, proportional to $\partial_yv_z+\partial_zv_y$. The former is generally found to be more than one order of magnitude larger than the latter and is largely confined at the fluid interfaces of the external and internal droplets (see Fig.\ref{fig12}). Interestingly, the stress tensor exhibits a self-similar pattern, an indication that the coupling between fluid flow and interface deformations weakly depends on the number of internal droplets. These results suggest that the interfacial stress can describe important aspects of the dynamics of the emulsion, such as periodic motion and interface deformations.

\section{Conclusions}

To summarize, we have investigated the dynamics of a 2D multi-core multiple emulsion under an imposed symmetric shear flow. We have kept the shear rate low enough to avoid the emulsion rupture but sufficiently intense to observe sizeable shape deformations. The physics of the steady states is crucially influenced by the shear rate and by the motion of the internal droplets. In the double emulsion, for instance, the latter undergoes shape deformations much weaker than the external one, which generally elongates until it attaines an elliptical shape aligned with the shear flow, a behavior similar to that observed in an isolated single fluid droplet.

On the other hand, higher complex systems exhibit novel non-equilibrium steady states. If, for instance, two smaller droplets are dispersed in a large one, an oscillatory steady state is produced, in which the internal droplets periodically rotate around the center of mass of the latter. This dynamics is rather robust, since it occurs regardless of the initial droplet arrangement and at higher shear rates. In addition, for moderate values of the shear rate, marked local deformations are produced at the interface of the extenal droplets of a multiple emulsion, due to the dynamic interaction with the internal ones. This result suggests that, alongside the Taylor parameter $D$ (a global quantity usually assumed to describe droplet deformations at low shear rates), futher quantities, such as the stress tensor, may be necessary to accurately capture these local interfacial bendings.

Our analysis shows that the dynamics of multiple emulsions shows non-trivial qualitatively new phenomena as compared to the case of double emulsions, thereby raising a number of open questions.

Is there, for example, a feasible strategy favouring an alternative dynamic behavior of the internal droplets, keeping the shear rate at low/moderate values? A potential route worth exploring would be to increase the volume fraction of the internal droplets up to the close-packing limit to possibly trigger a chaotic-like dynamics. This point will be investigated in a future work.

Furthermore, a more careful control of the droplet deformation is experimentally achieved by either gelling or hardening the state of the fluid in the layer, an effect that could be modeled by releasing the approximation of single viscosity adopted for both fluid components. Deformations are also notably affected by the physics of the droplet-droplet repulsive interactions, whose structure is determined by the nature of surfactant. A weak surfactant, for example, may favour a partial droplet merging, an effect that could in principle be controlled by properly tuning the strength $\epsilon$ of the repulsive term. Even more intriguing would be the study of the effect produced by the simultaneous presence of different surfactants, whose physics can be modeled by allowing each phase field its own repulsive strength $\epsilon(i)$.

Although still preliminary, we hope that our results may stimulate further experiments on multi-core emulsions in microfluidic devices, of potential interest in applications relevant to drug delivery or in food processing.

\section*{Appendix}

In Table \ref{tab:1} we provide an approximate mapping between simulation parameters and real units. Parameters have been chosen in agreement with values reported in previous simulation works, such as in \cite{utada2005monodisperse,marenduzzo1,marenduzzo2}. Realistic values can be obtained by fixing the lenght scale, the time scale and the force scale as: $L=1\mu$$m$, $T=10\mu$$s$ and $F=100$n$N$ (in simulations units these scales are all equal to one). 
\begin{table}[h!]
  \begin{center}
\caption{Typical values of the physical quantities used in the simulations.}\label{tab:1}
\vspace{0.2cm}
  \begin{tabular}{|c|c|c|}
	        \hline
                 {Model parameters} & {Simulation values} &  {Physical values} \\
                \hline
                Inner dropler radius, $R_{in}$ &   $10$  & $10$ $\mu$$m$ \\ 
                \hline
                Outer droplet radius, $R_{out}$ &   $50$  & $50$ $\mu$$m$ \\ 
                \hline
                Fluid viscosity, $\eta$ & $5/3$  & $\sim 10^{-2}$ $Pa\cdot s$ \\
                \hline
                Surface tension, $\sigma$ & $0.08$ & $\sim 1$m$N/m$ \\
                \hline
                Shear rate, $\dot{\gamma}$ & $10^{-4}$ & $\sim 0.02/s$ \\
                \hline
  \end{tabular}   
 \end{center}
\end{table}

\section*{Acknowledgments}
A. T., M. L., A. M., and S. S. acknowledge funding from the European Research Council under the European Union's Horizon 2020 Framework
Programme (No. FP/2014-2020) ERC Grant Agreement No.739964 (COPMAT).

\bibliographystyle{apsrev4-1}
\bibliography{bibliography}

\begin{thebibliography}{66}%
\makeatletter
\providecommand \@ifxundefined [1]{%
 \@ifx{#1\undefined}
}%
\providecommand \@ifnum [1]{%
 \ifnum #1\expandafter \@firstoftwo
 \else \expandafter \@secondoftwo
 \fi
}%
\providecommand \@ifx [1]{%
 \ifx #1\expandafter \@firstoftwo
 \else \expandafter \@secondoftwo
 \fi
}%
\providecommand \natexlab [1]{#1}%
\providecommand \enquote  [1]{``#1''}%
\providecommand \bibnamefont  [1]{#1}%
\providecommand \bibfnamefont [1]{#1}%
\providecommand \citenamefont [1]{#1}%
\providecommand \href@noop [0]{\@secondoftwo}%
\providecommand \href [0]{\begingroup \@sanitize@url \@href}%
\providecommand \@href[1]{\@@startlink{#1}\@@href}%
\providecommand \@@href[1]{\endgroup#1\@@endlink}%
\providecommand \@sanitize@url [0]{\catcode `\\12\catcode `\$12\catcode
  `\&12\catcode `\#12\catcode `\^12\catcode `\_12\catcode `\%12\relax}%
\providecommand \@@startlink[1]{}%
\providecommand \@@endlink[0]{}%
\providecommand \url  [0]{\begingroup\@sanitize@url \@url }%
\providecommand \@url [1]{\endgroup\@href {#1}{\urlprefix }}%
\providecommand \urlprefix  [0]{URL }%
\providecommand \Eprint [0]{\href }%
\providecommand \doibase [0]{http://dx.doi.org/}%
\providecommand \selectlanguage [0]{\@gobble}%
\providecommand \bibinfo  [0]{\@secondoftwo}%
\providecommand \bibfield  [0]{\@secondoftwo}%
\providecommand \translation [1]{[#1]}%
\providecommand \BibitemOpen [0]{}%
\providecommand \bibitemStop [0]{}%
\providecommand \bibitemNoStop [0]{.\EOS\space}%
\providecommand \EOS [0]{\spacefactor3000\relax}%
\providecommand \BibitemShut  [1]{\csname bibitem#1\endcsname}%
\let\auto@bib@innerbib\@empty
\bibitem [{\citenamefont {Lissant}(1974)}]{lissant1974emulsions}%
  \BibitemOpen
  \bibfield  {author} {\bibinfo {author} {\bibfnamefont {K.~J.}\ \bibnamefont
  {Lissant}},\ }\href@noop {} {\emph {\bibinfo {title} {Emulsions and
  Technology, vol. 6}}}\ (\bibinfo  {publisher} {Marcel Dekker, New York},\
  \bibinfo {year} {1974})\BibitemShut {NoStop}%
\bibitem [{\citenamefont {Kahn}\ \emph {et~al.}(2006)\citenamefont {Kahn},
  \citenamefont {Talegaonkar}, \citenamefont {Iqbal}, \citenamefont {Ahmend},\
  and\ \citenamefont {Khar}}]{kahn2006}%
  \BibitemOpen
  \bibfield  {author} {\bibinfo {author} {\bibfnamefont {A.~Y.}\ \bibnamefont
  {Kahn}}, \bibinfo {author} {\bibfnamefont {S.}~\bibnamefont {Talegaonkar}},
  \bibinfo {author} {\bibfnamefont {Z.}~\bibnamefont {Iqbal}}, \bibinfo
  {author} {\bibfnamefont {F.~J.}\ \bibnamefont {Ahmend}}, \ and\ \bibinfo
  {author} {\bibfnamefont {R.~K.}\ \bibnamefont {Khar}},\ }\href@noop {}
  {\bibfield  {journal} {\bibinfo  {journal} {Current Drug Delivery}\ }\textbf
  {\bibinfo {volume} {3}},\ \bibinfo {pages} {4} (\bibinfo {year}
  {2006})}\BibitemShut {NoStop}%
\bibitem [{\citenamefont {Datta}\ \emph {et~al.}(2014)\citenamefont {Datta},
  \citenamefont {Abbaspourrad}, \citenamefont {Amstad}, \citenamefont {Fan},
  \citenamefont {Kim}, \citenamefont {Romanowsky}, \citenamefont {Shum},
  \citenamefont {Sun}, \citenamefont {Utada}, \citenamefont {Windbergs},
  \citenamefont {Zhou},\ and\ \citenamefont {Weitz}}]{datta2014}%
  \BibitemOpen
  \bibfield  {author} {\bibinfo {author} {\bibfnamefont {S.}~\bibnamefont
  {Datta}}, \bibinfo {author} {\bibfnamefont {A.}~\bibnamefont {Abbaspourrad}},
  \bibinfo {author} {\bibfnamefont {E.}~\bibnamefont {Amstad}}, \bibinfo
  {author} {\bibfnamefont {J.}~\bibnamefont {Fan}}, \bibinfo {author}
  {\bibfnamefont {S.}~\bibnamefont {Kim}}, \bibinfo {author} {\bibfnamefont
  {M.}~\bibnamefont {Romanowsky}}, \bibinfo {author} {\bibfnamefont
  {H.}~\bibnamefont {Shum}}, \bibinfo {author} {\bibfnamefont {B.}~\bibnamefont
  {Sun}}, \bibinfo {author} {\bibfnamefont {A.}~\bibnamefont {Utada}}, \bibinfo
  {author} {\bibfnamefont {M.}~\bibnamefont {Windbergs}}, \bibinfo {author}
  {\bibfnamefont {S.}~\bibnamefont {Zhou}}, \ and\ \bibinfo {author}
  {\bibfnamefont {D.}~\bibnamefont {Weitz}},\ }\href@noop {} {\bibfield
  {journal} {\bibinfo  {journal} {Adv. Mater.}\ }\textbf {\bibinfo {volume}
  {26}},\ \bibinfo {pages} {2205} (\bibinfo {year} {2014})}\BibitemShut
  {NoStop}%
\bibitem [{\citenamefont {Vladisavljevic}\ \emph {et~al.}(2017)\citenamefont
  {Vladisavljevic}, \citenamefont {Al~Nuumani},\ and\ \citenamefont
  {Nabavi}}]{vladi2017}%
  \BibitemOpen
  \bibfield  {author} {\bibinfo {author} {\bibfnamefont {G.~T.}\ \bibnamefont
  {Vladisavljevic}}, \bibinfo {author} {\bibfnamefont {R.}~\bibnamefont
  {Al~Nuumani}}, \ and\ \bibinfo {author} {\bibfnamefont {S.~A.}\ \bibnamefont
  {Nabavi}},\ }\href@noop {} {\bibfield  {journal} {\bibinfo  {journal}
  {Micromachines}\ }\textbf {\bibinfo {volume} {8}},\ \bibinfo {pages} {75}
  (\bibinfo {year} {2017})}\BibitemShut {NoStop}%
\bibitem [{\citenamefont {Chu}\ \emph {et~al.}(2007)\citenamefont {Chu},
  \citenamefont {Utada}, \citenamefont {Shah}, \citenamefont {Kim},\ and\
  \citenamefont {Weitz}}]{weitz}%
  \BibitemOpen
  \bibfield  {author} {\bibinfo {author} {\bibfnamefont {L.}~\bibnamefont
  {Chu}}, \bibinfo {author} {\bibfnamefont {A.~S.}\ \bibnamefont {Utada}},
  \bibinfo {author} {\bibfnamefont {R.~K.}\ \bibnamefont {Shah}}, \bibinfo
  {author} {\bibfnamefont {J.~W.}\ \bibnamefont {Kim}}, \ and\ \bibinfo
  {author} {\bibfnamefont {D.~A.}\ \bibnamefont {Weitz}},\ }\href@noop {}
  {\bibfield  {journal} {\bibinfo  {journal} {Angew. Chem. Int. Ed. Engl.}\
  }\textbf {\bibinfo {volume} {46}},\ \bibinfo {pages} {8970} (\bibinfo {year}
  {2007})}\BibitemShut {NoStop}%
\bibitem [{\citenamefont {Xu}\ \emph {et~al.}(2006)\citenamefont {Xu},
  \citenamefont {Li}, \citenamefont {Tan}, \citenamefont {Wang},\ and\
  \citenamefont {Luo}}]{xu2006}%
  \BibitemOpen
  \bibfield  {author} {\bibinfo {author} {\bibfnamefont {J.~H.}\ \bibnamefont
  {Xu}}, \bibinfo {author} {\bibfnamefont {S.~W.}\ \bibnamefont {Li}}, \bibinfo
  {author} {\bibfnamefont {J.}~\bibnamefont {Tan}}, \bibinfo {author}
  {\bibfnamefont {Y.~J.}\ \bibnamefont {Wang}}, \ and\ \bibinfo {author}
  {\bibfnamefont {G.~S.}\ \bibnamefont {Luo}},\ }\href@noop {} {\bibfield
  {journal} {\bibinfo  {journal} {Langmuir}\ }\textbf {\bibinfo {volume}
  {22}},\ \bibinfo {pages} {7943} (\bibinfo {year} {2006})}\BibitemShut
  {NoStop}%
\bibitem [{\citenamefont {Santos}\ \emph {et~al.}(2016)\citenamefont {Santos},
  \citenamefont {Trujillo-Cayado}, \citenamefont {Calero}, \citenamefont
  {Alfaro},\ and\ \citenamefont {Munoz}}]{santos2016}%
  \BibitemOpen
  \bibfield  {author} {\bibinfo {author} {\bibfnamefont {J.}~\bibnamefont
  {Santos}}, \bibinfo {author} {\bibfnamefont {L.}~\bibnamefont
  {Trujillo-Cayado}}, \bibinfo {author} {\bibfnamefont {N.}~\bibnamefont
  {Calero}}, \bibinfo {author} {\bibfnamefont {M.}~\bibnamefont {Alfaro}}, \
  and\ \bibinfo {author} {\bibfnamefont {J.}~\bibnamefont {Munoz}},\
  }\href@noop {} {\bibfield  {journal} {\bibinfo  {journal} {J. Ind. Eng.
  Chem.}\ }\textbf {\bibinfo {volume} {36}},\ \bibinfo {pages} {90} (\bibinfo
  {year} {2016})}\BibitemShut {NoStop}%
\bibitem [{\citenamefont {Utada}\ \emph {et~al.}(2005)\citenamefont {Utada},
  \citenamefont {Lorenceau}, \citenamefont {Link}, \citenamefont {Kaplan},
  \citenamefont {Stone},\ and\ \citenamefont {Weitz}}]{utada2005monodisperse}%
  \BibitemOpen
  \bibfield  {author} {\bibinfo {author} {\bibfnamefont {A.}~\bibnamefont
  {Utada}}, \bibinfo {author} {\bibfnamefont {E.}~\bibnamefont {Lorenceau}},
  \bibinfo {author} {\bibfnamefont {D.}~\bibnamefont {Link}}, \bibinfo {author}
  {\bibfnamefont {P.}~\bibnamefont {Kaplan}}, \bibinfo {author} {\bibfnamefont
  {H.~A.}\ \bibnamefont {Stone}}, \ and\ \bibinfo {author} {\bibfnamefont
  {D.}~\bibnamefont {Weitz}},\ }\href@noop {} {\bibfield  {journal} {\bibinfo
  {journal} {Science}\ }\textbf {\bibinfo {volume} {308}},\ \bibinfo {pages}
  {537} (\bibinfo {year} {2005})}\BibitemShut {NoStop}%
\bibitem [{\citenamefont {Abate}\ and\ \citenamefont
  {Weitz}(2009)}]{abate2009}%
  \BibitemOpen
  \bibfield  {author} {\bibinfo {author} {\bibfnamefont {A.~R.}\ \bibnamefont
  {Abate}}\ and\ \bibinfo {author} {\bibfnamefont {D.~A.}\ \bibnamefont
  {Weitz}},\ }\href@noop {} {\bibfield  {journal} {\bibinfo  {journal} {Small}\
  }\textbf {\bibinfo {volume} {5}},\ \bibinfo {pages} {2030} (\bibinfo {year}
  {2009})}\BibitemShut {NoStop}%
\bibitem [{\citenamefont {Zarzar}\ \emph {et~al.}(2015)\citenamefont {Zarzar},
  \citenamefont {Sresht}, \citenamefont {Sletten}, \citenamefont {Kalow},
  \citenamefont {Blankschtein},\ and\ \citenamefont {Swager}}]{zarzar2015}%
  \BibitemOpen
  \bibfield  {author} {\bibinfo {author} {\bibfnamefont {L.}~\bibnamefont
  {Zarzar}}, \bibinfo {author} {\bibfnamefont {V.}~\bibnamefont {Sresht}},
  \bibinfo {author} {\bibfnamefont {E.}~\bibnamefont {Sletten}}, \bibinfo
  {author} {\bibfnamefont {J.}~\bibnamefont {Kalow}}, \bibinfo {author}
  {\bibfnamefont {D.}~\bibnamefont {Blankschtein}}, \ and\ \bibinfo {author}
  {\bibfnamefont {T.}~\bibnamefont {Swager}},\ }\href@noop {} {\bibfield
  {journal} {\bibinfo  {journal} {Nature}\ }\textbf {\bibinfo {volume} {518}},\
  \bibinfo {pages} {520} (\bibinfo {year} {2015})}\BibitemShut {NoStop}%
\bibitem [{\citenamefont {Cohen}\ \emph {et~al.}(1991)\citenamefont {Cohen},
  \citenamefont {Yoshioka}, \citenamefont {Lucarelli}, \citenamefont {Hwang},\
  and\ \citenamefont {Langer}}]{cohen1991controlled}%
  \BibitemOpen
  \bibfield  {author} {\bibinfo {author} {\bibfnamefont {S.}~\bibnamefont
  {Cohen}}, \bibinfo {author} {\bibfnamefont {T.}~\bibnamefont {Yoshioka}},
  \bibinfo {author} {\bibfnamefont {M.}~\bibnamefont {Lucarelli}}, \bibinfo
  {author} {\bibfnamefont {L.~H.}\ \bibnamefont {Hwang}}, \ and\ \bibinfo
  {author} {\bibfnamefont {R.}~\bibnamefont {Langer}},\ }\href@noop {}
  {\bibfield  {journal} {\bibinfo  {journal} {Pharmaceutical research}\
  }\textbf {\bibinfo {volume} {8}},\ \bibinfo {pages} {713} (\bibinfo {year}
  {1991})}\BibitemShut {NoStop}%
\bibitem [{\citenamefont {Laugel}\ \emph {et~al.}(2000)\citenamefont {Laugel},
  \citenamefont {Rafidison}, \citenamefont {Potard}, \citenamefont
  {Aguadisch},\ and\ \citenamefont {Baillet}}]{laugel2000modulated}%
  \BibitemOpen
  \bibfield  {author} {\bibinfo {author} {\bibfnamefont {C.}~\bibnamefont
  {Laugel}}, \bibinfo {author} {\bibfnamefont {P.}~\bibnamefont {Rafidison}},
  \bibinfo {author} {\bibfnamefont {G.}~\bibnamefont {Potard}}, \bibinfo
  {author} {\bibfnamefont {L.}~\bibnamefont {Aguadisch}}, \ and\ \bibinfo
  {author} {\bibfnamefont {A.}~\bibnamefont {Baillet}},\ }\href@noop {}
  {\bibfield  {journal} {\bibinfo  {journal} {Journal of controlled release}\
  }\textbf {\bibinfo {volume} {63}},\ \bibinfo {pages} {7} (\bibinfo {year}
  {2000})}\BibitemShut {NoStop}%
\bibitem [{\citenamefont {Cortesi}\ \emph {et~al.}(2002)\citenamefont
  {Cortesi}, \citenamefont {Esposito}, \citenamefont {Luca},\ and\
  \citenamefont {Nastruzzi}}]{cortesi2002production}%
  \BibitemOpen
  \bibfield  {author} {\bibinfo {author} {\bibfnamefont {R.}~\bibnamefont
  {Cortesi}}, \bibinfo {author} {\bibfnamefont {E.}~\bibnamefont {Esposito}},
  \bibinfo {author} {\bibfnamefont {G.}~\bibnamefont {Luca}}, \ and\ \bibinfo
  {author} {\bibfnamefont {C.}~\bibnamefont {Nastruzzi}},\ }\href@noop {}
  {\bibfield  {journal} {\bibinfo  {journal} {Biomaterials}\ }\textbf {\bibinfo
  {volume} {23}},\ \bibinfo {pages} {2283} (\bibinfo {year}
  {2002})}\BibitemShut {NoStop}%
\bibitem [{\citenamefont {Lamprecht}\ \emph {et~al.}(2004)\citenamefont
  {Lamprecht}, \citenamefont {Yamamoto}, \citenamefont {Takeuchi},\ and\
  \citenamefont {Kawashima}}]{lamprecht2004ph}%
  \BibitemOpen
  \bibfield  {author} {\bibinfo {author} {\bibfnamefont {A.}~\bibnamefont
  {Lamprecht}}, \bibinfo {author} {\bibfnamefont {H.}~\bibnamefont {Yamamoto}},
  \bibinfo {author} {\bibfnamefont {H.}~\bibnamefont {Takeuchi}}, \ and\
  \bibinfo {author} {\bibfnamefont {Y.}~\bibnamefont {Kawashima}},\ }\href@noop
  {} {\bibfield  {journal} {\bibinfo  {journal} {Journal of Controlled
  Release}\ }\textbf {\bibinfo {volume} {98}},\ \bibinfo {pages} {1} (\bibinfo
  {year} {2004})}\BibitemShut {NoStop}%
\bibitem [{\citenamefont {Kim}\ and\ \citenamefont
  {Park}(2004)}]{kim2004comparative}%
  \BibitemOpen
  \bibfield  {author} {\bibinfo {author} {\bibfnamefont {H.~K.}\ \bibnamefont
  {Kim}}\ and\ \bibinfo {author} {\bibfnamefont {T.~G.}\ \bibnamefont {Park}},\
  }\href@noop {} {\bibfield  {journal} {\bibinfo  {journal} {Journal of
  Controlled Release}\ }\textbf {\bibinfo {volume} {98}},\ \bibinfo {pages}
  {115} (\bibinfo {year} {2004})}\BibitemShut {NoStop}%
\bibitem [{\citenamefont {Chen}\ \emph {et~al.}(2008)\citenamefont {Chen},
  \citenamefont {Zhao}, \citenamefont {Song},\ and\ \citenamefont
  {Jiang}}]{chen2008one}%
  \BibitemOpen
  \bibfield  {author} {\bibinfo {author} {\bibfnamefont {H.}~\bibnamefont
  {Chen}}, \bibinfo {author} {\bibfnamefont {Y.}~\bibnamefont {Zhao}}, \bibinfo
  {author} {\bibfnamefont {Y.}~\bibnamefont {Song}}, \ and\ \bibinfo {author}
  {\bibfnamefont {L.}~\bibnamefont {Jiang}},\ }\href@noop {} {\bibfield
  {journal} {\bibinfo  {journal} {Journal of the American Chemical Society}\
  }\textbf {\bibinfo {volume} {130}},\ \bibinfo {pages} {7800} (\bibinfo {year}
  {2008})}\BibitemShut {NoStop}%
\bibitem [{\citenamefont {Lahann}(2011)}]{lahann2011recent}%
  \BibitemOpen
  \bibfield  {author} {\bibinfo {author} {\bibfnamefont {J.}~\bibnamefont
  {Lahann}},\ }\href@noop {} {\bibfield  {journal} {\bibinfo  {journal}
  {Small}\ }\textbf {\bibinfo {volume} {7}},\ \bibinfo {pages} {1149} (\bibinfo
  {year} {2011})}\BibitemShut {NoStop}%
\bibitem [{\citenamefont {Zhao}(2013)}]{zhao2013multiphase}%
  \BibitemOpen
  \bibfield  {author} {\bibinfo {author} {\bibfnamefont {C.-X.}\ \bibnamefont
  {Zhao}},\ }\href@noop {} {\bibfield  {journal} {\bibinfo  {journal} {Advanced
  drug delivery reviews}\ }\textbf {\bibinfo {volume} {65}},\ \bibinfo {pages}
  {1420} (\bibinfo {year} {2013})}\BibitemShut {NoStop}%
\bibitem [{\citenamefont {Zhang}\ \emph {et~al.}(2013)\citenamefont {Zhang},
  \citenamefont {Zhao}, \citenamefont {Rao}, \citenamefont {Snyder},
  \citenamefont {Choi}, \citenamefont {Wang}, \citenamefont {Khan},
  \citenamefont {Saleh}, \citenamefont {Mohler}, \citenamefont {Yu} \emph
  {et~al.}}]{zhang2013novel}%
  \BibitemOpen
  \bibfield  {author} {\bibinfo {author} {\bibfnamefont {W.}~\bibnamefont
  {Zhang}}, \bibinfo {author} {\bibfnamefont {S.}~\bibnamefont {Zhao}},
  \bibinfo {author} {\bibfnamefont {W.}~\bibnamefont {Rao}}, \bibinfo {author}
  {\bibfnamefont {J.}~\bibnamefont {Snyder}}, \bibinfo {author} {\bibfnamefont
  {J.~K.}\ \bibnamefont {Choi}}, \bibinfo {author} {\bibfnamefont
  {J.}~\bibnamefont {Wang}}, \bibinfo {author} {\bibfnamefont {I.~A.}\
  \bibnamefont {Khan}}, \bibinfo {author} {\bibfnamefont {N.~B.}\ \bibnamefont
  {Saleh}}, \bibinfo {author} {\bibfnamefont {P.~J.}\ \bibnamefont {Mohler}},
  \bibinfo {author} {\bibfnamefont {J.}~\bibnamefont {Yu}},  \emph {et~al.},\
  }\href@noop {} {\bibfield  {journal} {\bibinfo  {journal} {Journal of
  materials chemistry B}\ }\textbf {\bibinfo {volume} {1}},\ \bibinfo {pages}
  {1002} (\bibinfo {year} {2013})}\BibitemShut {NoStop}%
\bibitem [{\citenamefont {Li}\ and\ \citenamefont
  {Shrier}(1972)}]{li1972liquid}%
  \BibitemOpen
  \bibfield  {author} {\bibinfo {author} {\bibfnamefont {N.~N.}\ \bibnamefont
  {Li}}\ and\ \bibinfo {author} {\bibfnamefont {A.}~\bibnamefont {Shrier}},\
  }\href@noop {} {\bibfield  {journal} {\bibinfo  {journal} {Recent
  Developments in Separation Science}\ }\textbf {\bibinfo {volume} {1}},\
  \bibinfo {pages} {163} (\bibinfo {year} {1972})}\BibitemShut {NoStop}%
\bibitem [{\citenamefont {Muguet}\ \emph {et~al.}(2001)\citenamefont {Muguet},
  \citenamefont {Seiller}, \citenamefont {Barratt}, \citenamefont {Ozer},
  \citenamefont {Marty},\ and\ \citenamefont
  {Grossiord}}]{muguet2001formulation}%
  \BibitemOpen
  \bibfield  {author} {\bibinfo {author} {\bibfnamefont {V.}~\bibnamefont
  {Muguet}}, \bibinfo {author} {\bibfnamefont {M.}~\bibnamefont {Seiller}},
  \bibinfo {author} {\bibfnamefont {G.}~\bibnamefont {Barratt}}, \bibinfo
  {author} {\bibfnamefont {O.}~\bibnamefont {Ozer}}, \bibinfo {author}
  {\bibfnamefont {J.}~\bibnamefont {Marty}}, \ and\ \bibinfo {author}
  {\bibfnamefont {J.}~\bibnamefont {Grossiord}},\ }\href@noop {} {\bibfield
  {journal} {\bibinfo  {journal} {Journal of controlled release}\ }\textbf
  {\bibinfo {volume} {70}},\ \bibinfo {pages} {37} (\bibinfo {year}
  {2001})}\BibitemShut {NoStop}%
\bibitem [{\citenamefont {Lee}\ \emph {et~al.}(2001)\citenamefont {Lee},
  \citenamefont {Oh}, \citenamefont {Moon},\ and\ \citenamefont
  {Bae}}]{lee2001preparation}%
  \BibitemOpen
  \bibfield  {author} {\bibinfo {author} {\bibfnamefont {M.-H.}\ \bibnamefont
  {Lee}}, \bibinfo {author} {\bibfnamefont {S.-G.}\ \bibnamefont {Oh}},
  \bibinfo {author} {\bibfnamefont {S.-K.}\ \bibnamefont {Moon}}, \ and\
  \bibinfo {author} {\bibfnamefont {S.-Y.}\ \bibnamefont {Bae}},\ }\href@noop
  {} {\bibfield  {journal} {\bibinfo  {journal} {Journal of colloid and
  interface science}\ }\textbf {\bibinfo {volume} {240}},\ \bibinfo {pages}
  {83} (\bibinfo {year} {2001})}\BibitemShut {NoStop}%
\bibitem [{\citenamefont {Lee}\ \emph {et~al.}(2002)\citenamefont {Lee},
  \citenamefont {Goh}, \citenamefont {Kim}, \citenamefont {Kim}, \citenamefont
  {Kang}, \citenamefont {Suh},\ and\ \citenamefont {Kim}}]{lee2002effective}%
  \BibitemOpen
  \bibfield  {author} {\bibinfo {author} {\bibfnamefont {D.-H.}\ \bibnamefont
  {Lee}}, \bibinfo {author} {\bibfnamefont {Y.-M.}\ \bibnamefont {Goh}},
  \bibinfo {author} {\bibfnamefont {J.-S.}\ \bibnamefont {Kim}}, \bibinfo
  {author} {\bibfnamefont {H.-K.}\ \bibnamefont {Kim}}, \bibinfo {author}
  {\bibfnamefont {H.-H.}\ \bibnamefont {Kang}}, \bibinfo {author}
  {\bibfnamefont {K.-D.}\ \bibnamefont {Suh}}, \ and\ \bibinfo {author}
  {\bibfnamefont {J.-W.}\ \bibnamefont {Kim}},\ }\href@noop {} {\bibfield
  {journal} {\bibinfo  {journal} {Journal of dispersion science and
  technology}\ }\textbf {\bibinfo {volume} {23}},\ \bibinfo {pages} {491}
  (\bibinfo {year} {2002})}\BibitemShut {NoStop}%
\bibitem [{\citenamefont {Edris}\ and\ \citenamefont
  {Bergnst{\aa}hl}(2001)}]{edris2001encapsulation}%
  \BibitemOpen
  \bibfield  {author} {\bibinfo {author} {\bibfnamefont {A.}~\bibnamefont
  {Edris}}\ and\ \bibinfo {author} {\bibfnamefont {B.}~\bibnamefont
  {Bergnst{\aa}hl}},\ }\href@noop {} {\bibfield  {journal} {\bibinfo  {journal}
  {Food/Nahrung}\ }\textbf {\bibinfo {volume} {45}},\ \bibinfo {pages} {133}
  (\bibinfo {year} {2001})}\BibitemShut {NoStop}%
\bibitem [{\citenamefont {Benichou}\ \emph {et~al.}(2002)\citenamefont
  {Benichou}, \citenamefont {Aserin},\ and\ \citenamefont
  {Garti}}]{benichou2002double}%
  \BibitemOpen
  \bibfield  {author} {\bibinfo {author} {\bibfnamefont {A.}~\bibnamefont
  {Benichou}}, \bibinfo {author} {\bibfnamefont {A.}~\bibnamefont {Aserin}}, \
  and\ \bibinfo {author} {\bibfnamefont {N.}~\bibnamefont {Garti}},\
  }\href@noop {} {\bibfield  {journal} {\bibinfo  {journal} {Polymers for
  Advanced Technologies}\ }\textbf {\bibinfo {volume} {13}},\ \bibinfo {pages}
  {1019} (\bibinfo {year} {2002})}\BibitemShut {NoStop}%
\bibitem [{\citenamefont {Omi}\ \emph {et~al.}(2003)\citenamefont {Omi},
  \citenamefont {Katami}, \citenamefont {Taguchi}, \citenamefont {Kaneko},\
  and\ \citenamefont {Iso}}]{omi2003}%
  \BibitemOpen
  \bibfield  {author} {\bibinfo {author} {\bibfnamefont {S.}~\bibnamefont
  {Omi}}, \bibinfo {author} {\bibfnamefont {K.}~\bibnamefont {Katami}},
  \bibinfo {author} {\bibfnamefont {T.}~\bibnamefont {Taguchi}}, \bibinfo
  {author} {\bibfnamefont {K.}~\bibnamefont {Kaneko}}, \ and\ \bibinfo {author}
  {\bibfnamefont {M.}~\bibnamefont {Iso}},\ }\href@noop {} {\bibfield
  {journal} {\bibinfo  {journal} {J. Appl. Polym. Sci}\ }\textbf {\bibinfo
  {volume} {57}},\ \bibinfo {pages} {1013} (\bibinfo {year}
  {2003})}\BibitemShut {NoStop}%
\bibitem [{\citenamefont {Chu}\ \emph {et~al.}(2003)\citenamefont {Chu},
  \citenamefont {Xie}, \citenamefont {Zhu}, \citenamefont {Chen}, \citenamefont
  {Yamaguchi},\ and\ \citenamefont {Nakao}}]{chu2003}%
  \BibitemOpen
  \bibfield  {author} {\bibinfo {author} {\bibfnamefont {L.~Y.}\ \bibnamefont
  {Chu}}, \bibinfo {author} {\bibfnamefont {R.}~\bibnamefont {Xie}}, \bibinfo
  {author} {\bibfnamefont {J.~H.}\ \bibnamefont {Zhu}}, \bibinfo {author}
  {\bibfnamefont {W.~M.}\ \bibnamefont {Chen}}, \bibinfo {author}
  {\bibfnamefont {T.}~\bibnamefont {Yamaguchi}}, \ and\ \bibinfo {author}
  {\bibfnamefont {S.~I.}\ \bibnamefont {Nakao}},\ }\href@noop {} {\bibfield
  {journal} {\bibinfo  {journal} {J. Colloid Interface Sci}\ }\textbf {\bibinfo
  {volume} {265}},\ \bibinfo {pages} {187} (\bibinfo {year}
  {2003})}\BibitemShut {NoStop}%
\bibitem [{\citenamefont {Alex}\ and\ \citenamefont
  {Bodmeier}(1990)}]{alex1990}%
  \BibitemOpen
  \bibfield  {author} {\bibinfo {author} {\bibfnamefont {R.}~\bibnamefont
  {Alex}}\ and\ \bibinfo {author} {\bibfnamefont {R.}~\bibnamefont
  {Bodmeier}},\ }\href@noop {} {\bibfield  {journal} {\bibinfo  {journal} {J.
  Microencapsul.}\ }\textbf {\bibinfo {volume} {7}},\ \bibinfo {pages} {347}
  (\bibinfo {year} {1990})}\BibitemShut {NoStop}%
\bibitem [{\citenamefont {Kim}\ and\ \citenamefont {Weitz}(2011)}]{kim2011one}%
  \BibitemOpen
  \bibfield  {author} {\bibinfo {author} {\bibfnamefont {S.-H.}\ \bibnamefont
  {Kim}}\ and\ \bibinfo {author} {\bibfnamefont {D.~A.}\ \bibnamefont
  {Weitz}},\ }\href@noop {} {\bibfield  {journal} {\bibinfo  {journal}
  {Angewandte Chemie International Edition}\ }\textbf {\bibinfo {volume}
  {50}},\ \bibinfo {pages} {8731} (\bibinfo {year} {2011})}\BibitemShut
  {NoStop}%
\bibitem [{\citenamefont {i~Solvas}\ and\ \citenamefont
  {DeMello}(2011)}]{i2011droplet}%
  \BibitemOpen
  \bibfield  {author} {\bibinfo {author} {\bibfnamefont {X.~C.}\ \bibnamefont
  {i~Solvas}}\ and\ \bibinfo {author} {\bibfnamefont {A.}~\bibnamefont
  {DeMello}},\ }\href@noop {} {\bibfield  {journal} {\bibinfo  {journal}
  {Chemical Communications}\ }\textbf {\bibinfo {volume} {47}},\ \bibinfo
  {pages} {1936} (\bibinfo {year} {2011})}\BibitemShut {NoStop}%
\bibitem [{\citenamefont {Saeki}\ \emph {et~al.}(2010)\citenamefont {Saeki},
  \citenamefont {Sugiura}, \citenamefont {Kanamori}, \citenamefont {Sato},\
  and\ \citenamefont {Ichikawa}}]{saeki2010}%
  \BibitemOpen
  \bibfield  {author} {\bibinfo {author} {\bibfnamefont {D.}~\bibnamefont
  {Saeki}}, \bibinfo {author} {\bibfnamefont {S.}~\bibnamefont {Sugiura}},
  \bibinfo {author} {\bibfnamefont {T.}~\bibnamefont {Kanamori}}, \bibinfo
  {author} {\bibfnamefont {S.}~\bibnamefont {Sato}}, \ and\ \bibinfo {author}
  {\bibfnamefont {S.}~\bibnamefont {Ichikawa}},\ }\href@noop {} {\bibfield
  {journal} {\bibinfo  {journal} {Lab Chip}\ }\textbf {\bibinfo {volume}
  {10}},\ \bibinfo {pages} {357} (\bibinfo {year} {2010})}\BibitemShut
  {NoStop}%
\bibitem [{\citenamefont {Chen}\ and\ \citenamefont {Shi}(2013)}]{chen1}%
  \BibitemOpen
  \bibfield  {author} {\bibinfo {author} {\bibfnamefont {X.}~\bibnamefont
  {Chen}, \bibfnamefont {Y.~Liu}}\ and\ \bibinfo {author} {\bibfnamefont
  {M.}~\bibnamefont {Shi}},\ }\href@noop {} {\bibfield  {journal} {\bibinfo
  {journal} {Appl. Phys. Lett.}\ }\textbf {\bibinfo {volume} {102}},\ \bibinfo
  {pages} {061609} (\bibinfo {year} {2013})}\BibitemShut {NoStop}%
\bibitem [{\citenamefont {Chen}\ \emph
  {et~al.}(2015{\natexlab{a}})\citenamefont {Chen}, \citenamefont {Liu},\ and\
  \citenamefont {Zhao}}]{chen2}%
  \BibitemOpen
  \bibfield  {author} {\bibinfo {author} {\bibfnamefont {Y.}~\bibnamefont
  {Chen}}, \bibinfo {author} {\bibfnamefont {X.}~\bibnamefont {Liu}}, \ and\
  \bibinfo {author} {\bibfnamefont {Y.}~\bibnamefont {Zhao}},\ }\href@noop {}
  {\bibfield  {journal} {\bibinfo  {journal} {Appl. Phys. Lett.}\ }\textbf
  {\bibinfo {volume} {106}},\ \bibinfo {pages} {141601} (\bibinfo {year}
  {2015}{\natexlab{a}})}\BibitemShut {NoStop}%
\bibitem [{\citenamefont {Chen}\ \emph
  {et~al.}(2015{\natexlab{b}})\citenamefont {Chen}, \citenamefont {Xiangdong},
  \citenamefont {Chengbin},\ and\ \citenamefont {Yuanjin}}]{chen3}%
  \BibitemOpen
  \bibfield  {author} {\bibinfo {author} {\bibfnamefont {Y.}~\bibnamefont
  {Chen}}, \bibinfo {author} {\bibfnamefont {L.}~\bibnamefont {Xiangdong}},
  \bibinfo {author} {\bibfnamefont {Z.}~\bibnamefont {Chengbin}}, \ and\
  \bibinfo {author} {\bibfnamefont {Z.}~\bibnamefont {Yuanjin}},\ }\href@noop
  {} {\bibfield  {journal} {\bibinfo  {journal} {Lab on a Chip}\ }\textbf
  {\bibinfo {volume} {15}},\ \bibinfo {pages} {1255} (\bibinfo {year}
  {2015}{\natexlab{b}})}\BibitemShut {NoStop}%
\bibitem [{\citenamefont {Stone}\ and\ \citenamefont {Leal}(1990)}]{stone}%
  \BibitemOpen
  \bibfield  {author} {\bibinfo {author} {\bibfnamefont {H.~A.}\ \bibnamefont
  {Stone}}\ and\ \bibinfo {author} {\bibfnamefont {L.~G.}\ \bibnamefont
  {Leal}},\ }\href@noop {} {\bibfield  {journal} {\bibinfo  {journal} {J.
  Fluid. Mech.}\ }\textbf {\bibinfo {volume} {211}},\ \bibinfo {pages} {123}
  (\bibinfo {year} {1990})}\BibitemShut {NoStop}%
\bibitem [{\citenamefont {Renardy}\ and\ \citenamefont
  {Cristini}(2001)}]{renardy}%
  \BibitemOpen
  \bibfield  {author} {\bibinfo {author} {\bibfnamefont {Y.~Y.}\ \bibnamefont
  {Renardy}}\ and\ \bibinfo {author} {\bibfnamefont {V.}~\bibnamefont
  {Cristini}},\ }\href@noop {} {\bibfield  {journal} {\bibinfo  {journal}
  {Physics of Fluids}\ }\textbf {\bibinfo {volume} {13}},\ \bibinfo {pages} {7}
  (\bibinfo {year} {2001})}\BibitemShut {NoStop}%
\bibitem [{\citenamefont {Afkami}\ \emph {et~al.}(2009)\citenamefont {Afkami},
  \citenamefont {Yue},\ and\ \citenamefont {Renardy}}]{afkhami2009}%
  \BibitemOpen
  \bibfield  {author} {\bibinfo {author} {\bibfnamefont {S.}~\bibnamefont
  {Afkami}}, \bibinfo {author} {\bibfnamefont {P.}~\bibnamefont {Yue}}, \ and\
  \bibinfo {author} {\bibfnamefont {Y.}~\bibnamefont {Renardy}},\ }\href@noop
  {} {\bibfield  {journal} {\bibinfo  {journal} {Physics of Fluids}\ }\textbf
  {\bibinfo {volume} {21}},\ \bibinfo {pages} {072106} (\bibinfo {year}
  {2009})}\BibitemShut {NoStop}%
\bibitem [{\citenamefont {Ha}\ and\ \citenamefont {Yang}(1999)}]{ha1999}%
  \BibitemOpen
  \bibfield  {author} {\bibinfo {author} {\bibfnamefont {J.~W.}\ \bibnamefont
  {Ha}}\ and\ \bibinfo {author} {\bibfnamefont {S.~M.}\ \bibnamefont {Yang}},\
  }\href@noop {} {\bibfield  {journal} {\bibinfo  {journal} {Physics of
  Fluids}\ }\textbf {\bibinfo {volume} {11}},\ \bibinfo {pages} {1029}
  (\bibinfo {year} {1999})}\BibitemShut {NoStop}%
\bibitem [{\citenamefont {Wang}\ \emph {et~al.}(2013)\citenamefont {Wang},
  \citenamefont {Liu}, \citenamefont {Han},\ and\ \citenamefont {Guan}}]{wang}%
  \BibitemOpen
  \bibfield  {author} {\bibinfo {author} {\bibfnamefont {J.}~\bibnamefont
  {Wang}}, \bibinfo {author} {\bibfnamefont {J.}~\bibnamefont {Liu}}, \bibinfo
  {author} {\bibfnamefont {J.}~\bibnamefont {Han}}, \ and\ \bibinfo {author}
  {\bibfnamefont {J.}~\bibnamefont {Guan}},\ }\href@noop {} {\bibfield
  {journal} {\bibinfo  {journal} {Phys. Rev. Lett.}\ }\textbf {\bibinfo
  {volume} {110}},\ \bibinfo {pages} {066001} (\bibinfo {year}
  {2013})}\BibitemShut {NoStop}%
\bibitem [{\citenamefont {Smith}\ \emph {et~al.}(2004)\citenamefont {Smith},
  \citenamefont {Ottino},\ and\ \citenamefont {Olvera de~la Cruz}}]{smith}%
  \BibitemOpen
  \bibfield  {author} {\bibinfo {author} {\bibfnamefont {K.~A.}\ \bibnamefont
  {Smith}}, \bibinfo {author} {\bibfnamefont {J.~M.}\ \bibnamefont {Ottino}}, \
  and\ \bibinfo {author} {\bibfnamefont {M.}~\bibnamefont {Olvera de~la
  Cruz}},\ }\href@noop {} {\bibfield  {journal} {\bibinfo  {journal} {Phys.
  Rev. Lett.}\ }\textbf {\bibinfo {volume} {93}},\ \bibinfo {pages} {204501}
  (\bibinfo {year} {2004})}\BibitemShut {NoStop}%
\bibitem [{\citenamefont {Foglino}\ \emph {et~al.}(2017)\citenamefont
  {Foglino}, \citenamefont {Morozov}, \citenamefont {Henrich},\ and\
  \citenamefont {Marenduzzo}}]{marenduzzo1}%
  \BibitemOpen
  \bibfield  {author} {\bibinfo {author} {\bibfnamefont {M.}~\bibnamefont
  {Foglino}}, \bibinfo {author} {\bibfnamefont {A.~N.}\ \bibnamefont
  {Morozov}}, \bibinfo {author} {\bibfnamefont {O.}~\bibnamefont {Henrich}}, \
  and\ \bibinfo {author} {\bibfnamefont {D.}~\bibnamefont {Marenduzzo}},\
  }\href@noop {} {\bibfield  {journal} {\bibinfo  {journal} {Phys. Rev. Lett.}\
  }\textbf {\bibinfo {volume} {119}},\ \bibinfo {pages} {208802} (\bibinfo
  {year} {2017})}\BibitemShut {NoStop}%
\bibitem [{\citenamefont {Foglino}\ \emph {et~al.}(2018)\citenamefont
  {Foglino}, \citenamefont {Morozov},\ and\ \citenamefont
  {Marenduzzo}}]{marenduzzo2}%
  \BibitemOpen
  \bibfield  {author} {\bibinfo {author} {\bibfnamefont {M.}~\bibnamefont
  {Foglino}}, \bibinfo {author} {\bibfnamefont {A.~N.}\ \bibnamefont
  {Morozov}}, \ and\ \bibinfo {author} {\bibfnamefont {D.}~\bibnamefont
  {Marenduzzo}},\ }\href@noop {} {\bibfield  {journal} {\bibinfo  {journal}
  {Soft Matter}\ }\textbf {\bibinfo {volume} {14}},\ \bibinfo {pages} {9361}
  (\bibinfo {year} {2018})}\BibitemShut {NoStop}%
\bibitem [{\citenamefont {De~Groot}\ and\ \citenamefont
  {Mazur}(1984)}]{degroot}%
  \BibitemOpen
  \bibfield  {author} {\bibinfo {author} {\bibfnamefont {S.~R.}\ \bibnamefont
  {De~Groot}}\ and\ \bibinfo {author} {\bibfnamefont {P.}~\bibnamefont
  {Mazur}},\ }\href@noop {} {\emph {\bibinfo {title} {Non-Equilibrium
  Thermodynamics}}}\ (\bibinfo  {publisher} {New York, NY, Dover},\ \bibinfo
  {year} {1984})\BibitemShut {NoStop}%
\bibitem [{\citenamefont {Mueller}\ \emph {et~al.}(2019)\citenamefont
  {Mueller}, \citenamefont {M.},\ and\ \citenamefont
  {Doostmohammadi}}]{yeomans}%
  \BibitemOpen
  \bibfield  {author} {\bibinfo {author} {\bibfnamefont {R.}~\bibnamefont
  {Mueller}}, \bibinfo {author} {\bibfnamefont {Y.~J.}\ \bibnamefont {M.}}, \
  and\ \bibinfo {author} {\bibfnamefont {A.}~\bibnamefont {Doostmohammadi}},\
  }\href@noop {} {\bibfield  {journal} {\bibinfo  {journal} {Phys. Rev. Lett.}\
  }\textbf {\bibinfo {volume} {122}},\ \bibinfo {pages} {048004} (\bibinfo
  {year} {2019})}\BibitemShut {NoStop}%
\bibitem [{\citenamefont {Kendon}\ \emph {et~al.}(2001)\citenamefont {Kendon},
  \citenamefont {Cates}, \citenamefont {Pagonabarraga}, \citenamefont
  {Desplat},\ and\ \citenamefont {Blandon}}]{cates1}%
  \BibitemOpen
  \bibfield  {author} {\bibinfo {author} {\bibfnamefont {V.~M.}\ \bibnamefont
  {Kendon}}, \bibinfo {author} {\bibfnamefont {M.~E.}\ \bibnamefont {Cates}},
  \bibinfo {author} {\bibfnamefont {I.}~\bibnamefont {Pagonabarraga}}, \bibinfo
  {author} {\bibfnamefont {J.~C.}\ \bibnamefont {Desplat}}, \ and\ \bibinfo
  {author} {\bibfnamefont {P.}~\bibnamefont {Blandon}},\ }\href@noop {}
  {\bibfield  {journal} {\bibinfo  {journal} {J. Fluid. Mech.}\ }\textbf
  {\bibinfo {volume} {440}},\ \bibinfo {pages} {147} (\bibinfo {year}
  {2001})}\BibitemShut {NoStop}%
\bibitem [{\citenamefont {Succi}(2018)}]{succi1}%
  \BibitemOpen
  \bibfield  {author} {\bibinfo {author} {\bibfnamefont {S.}~\bibnamefont
  {Succi}},\ }\href@noop {} {\emph {\bibinfo {title} {The Lattice Boltzmann
  Equation: For Complex States of Flowing Matter}}}\ (\bibinfo  {publisher}
  {Oxford University Press},\ \bibinfo {year} {2018})\BibitemShut {NoStop}%
\bibitem [{\citenamefont {Benzi}\ \emph {et~al.}(1992)\citenamefont {Benzi},
  \citenamefont {Succi},\ and\ \citenamefont {Vergassola}}]{succi2}%
  \BibitemOpen
  \bibfield  {author} {\bibinfo {author} {\bibfnamefont {R.}~\bibnamefont
  {Benzi}}, \bibinfo {author} {\bibfnamefont {S.}~\bibnamefont {Succi}}, \ and\
  \bibinfo {author} {\bibfnamefont {M.}~\bibnamefont {Vergassola}},\
  }\href@noop {} {\bibfield  {journal} {\bibinfo  {journal} {Phys. Rep.}\
  }\textbf {\bibinfo {volume} {222}},\ \bibinfo {pages} {145} (\bibinfo {year}
  {1992})}\BibitemShut {NoStop}%
\bibitem [{\citenamefont {Kr{\"u}ger}\ \emph {et~al.}(2017)\citenamefont
  {Kr{\"u}ger}, \citenamefont {Kusumaatmaja}, \citenamefont {Kuzmin},
  \citenamefont {Shardt}, \citenamefont {Silva},\ and\ \citenamefont
  {Viggen}}]{kruger}%
  \BibitemOpen
  \bibfield  {author} {\bibinfo {author} {\bibfnamefont {T.}~\bibnamefont
  {Kr{\"u}ger}}, \bibinfo {author} {\bibfnamefont {H.}~\bibnamefont
  {Kusumaatmaja}}, \bibinfo {author} {\bibfnamefont {A.}~\bibnamefont
  {Kuzmin}}, \bibinfo {author} {\bibfnamefont {O.}~\bibnamefont {Shardt}},
  \bibinfo {author} {\bibfnamefont {G.}~\bibnamefont {Silva}}, \ and\ \bibinfo
  {author} {\bibfnamefont {E.~M.}\ \bibnamefont {Viggen}},\ }\href@noop {}
  {\bibfield  {journal} {\bibinfo  {journal} {Springer International
  Publishing}\ }\textbf {\bibinfo {volume} {10}},\ \bibinfo {pages} {978}
  (\bibinfo {year} {2017})}\BibitemShut {NoStop}%
\bibitem [{\citenamefont {Swift}\ \emph {et~al.}(1996)\citenamefont {Swift},
  \citenamefont {Orlandini}, \citenamefont {Osborn},\ and\ \citenamefont
  {Yeomans}}]{yeomans2}%
  \BibitemOpen
  \bibfield  {author} {\bibinfo {author} {\bibfnamefont {M.~R.}\ \bibnamefont
  {Swift}}, \bibinfo {author} {\bibfnamefont {E.}~\bibnamefont {Orlandini}},
  \bibinfo {author} {\bibfnamefont {W.~R.}\ \bibnamefont {Osborn}}, \ and\
  \bibinfo {author} {\bibfnamefont {J.~M.}\ \bibnamefont {Yeomans}},\
  }\href@noop {} {\bibfield  {journal} {\bibinfo  {journal} {Phys. Rev. E}\
  }\textbf {\bibinfo {volume} {54}},\ \bibinfo {pages} {5041} (\bibinfo {year}
  {1996})}\BibitemShut {NoStop}%
\bibitem [{\citenamefont {Bernaschi}\ \emph {et~al.}(2019)\citenamefont
  {Bernaschi}, \citenamefont {Melchionna},\ and\ \citenamefont
  {Succi}}]{succi3}%
  \BibitemOpen
  \bibfield  {author} {\bibinfo {author} {\bibfnamefont {M.}~\bibnamefont
  {Bernaschi}}, \bibinfo {author} {\bibfnamefont {S.}~\bibnamefont
  {Melchionna}}, \ and\ \bibinfo {author} {\bibfnamefont {S.}~\bibnamefont
  {Succi}},\ }\href@noop {} {\bibfield  {journal} {\bibinfo  {journal} {Rev.
  Mod. Phys.}\ }\textbf {\bibinfo {volume} {91}},\ \bibinfo {pages} {025004}
  (\bibinfo {year} {2019})}\BibitemShut {NoStop}%
\bibitem [{\citenamefont {Montessori}\ \emph {et~al.}(2019)\citenamefont
  {Montessori}, \citenamefont {Lauricella}, \citenamefont {Tirelli},\ and\
  \citenamefont {Succi}}]{montessori}%
  \BibitemOpen
  \bibfield  {author} {\bibinfo {author} {\bibfnamefont {A.}~\bibnamefont
  {Montessori}}, \bibinfo {author} {\bibfnamefont {M.}~\bibnamefont
  {Lauricella}}, \bibinfo {author} {\bibfnamefont {N.}~\bibnamefont {Tirelli}},
  \ and\ \bibinfo {author} {\bibfnamefont {S.}~\bibnamefont {Succi}},\
  }\href@noop {} {\bibfield  {journal} {\bibinfo  {journal} {Journ. Fluid.
  Mech.}\ }\textbf {\bibinfo {volume} {872}},\ \bibinfo {pages} {327} (\bibinfo
  {year} {2019})}\BibitemShut {NoStop}%
\bibitem [{\citenamefont {Ansumali}\ and\ \citenamefont
  {Karlin}(2002)}]{karlin}%
  \BibitemOpen
  \bibfield  {author} {\bibinfo {author} {\bibfnamefont {S.}~\bibnamefont
  {Ansumali}}\ and\ \bibinfo {author} {\bibfnamefont {I.~V.}\ \bibnamefont
  {Karlin}},\ }\href@noop {} {\bibfield  {journal} {\bibinfo  {journal} {Phys.
  Rev. E}\ }\textbf {\bibinfo {volume} {66}},\ \bibinfo {pages} {026311}
  (\bibinfo {year} {2002})}\BibitemShut {NoStop}%
\bibitem [{\citenamefont {Ansumali}\ \emph {et~al.}(2007)\citenamefont
  {Ansumali}, \citenamefont {Karlin}, \citenamefont {Arcidiacono},
  \citenamefont {Abbas},\ and\ \citenamefont {Prasianakis}}]{ansumali}%
  \BibitemOpen
  \bibfield  {author} {\bibinfo {author} {\bibfnamefont {S.}~\bibnamefont
  {Ansumali}}, \bibinfo {author} {\bibfnamefont {I.~V.}\ \bibnamefont
  {Karlin}}, \bibinfo {author} {\bibfnamefont {S.}~\bibnamefont {Arcidiacono}},
  \bibinfo {author} {\bibfnamefont {S.}~\bibnamefont {Abbas}}, \ and\ \bibinfo
  {author} {\bibfnamefont {N.~I.}\ \bibnamefont {Prasianakis}},\ }\href@noop {}
  {\bibfield  {journal} {\bibinfo  {journal} {Phys. Rev. Lett.}\ }\textbf
  {\bibinfo {volume} {98}},\ \bibinfo {pages} {124502} (\bibinfo {year}
  {2007})}\BibitemShut {NoStop}%
\bibitem [{\citenamefont {Tiribocchi}\ \emph {et~al.}(2009)\citenamefont
  {Tiribocchi}, \citenamefont {Stella}, \citenamefont {Gonnella},\ and\
  \citenamefont {Lamura}}]{tiribocchi}%
  \BibitemOpen
  \bibfield  {author} {\bibinfo {author} {\bibfnamefont {A.}~\bibnamefont
  {Tiribocchi}}, \bibinfo {author} {\bibfnamefont {N.}~\bibnamefont {Stella}},
  \bibinfo {author} {\bibfnamefont {G.}~\bibnamefont {Gonnella}}, \ and\
  \bibinfo {author} {\bibfnamefont {A.}~\bibnamefont {Lamura}},\ }\href@noop {}
  {\bibfield  {journal} {\bibinfo  {journal} {Phys. Rev. E}\ }\textbf {\bibinfo
  {volume} {80}},\ \bibinfo {pages} {026701} (\bibinfo {year}
  {2009})}\BibitemShut {NoStop}%
\bibitem [{\citenamefont {Gonnella}\ \emph {et~al.}(2010)\citenamefont
  {Gonnella}, \citenamefont {Lamura}, \citenamefont {Piscitelli},\ and\
  \citenamefont {Tiribocchi}}]{tiribocchi2}%
  \BibitemOpen
  \bibfield  {author} {\bibinfo {author} {\bibfnamefont {G.}~\bibnamefont
  {Gonnella}}, \bibinfo {author} {\bibfnamefont {A.}~\bibnamefont {Lamura}},
  \bibinfo {author} {\bibfnamefont {A.}~\bibnamefont {Piscitelli}}, \ and\
  \bibinfo {author} {\bibfnamefont {A.}~\bibnamefont {Tiribocchi}},\
  }\href@noop {} {\bibfield  {journal} {\bibinfo  {journal} {Phys. Rev. E.}\
  }\textbf {\bibinfo {volume} {82}},\ \bibinfo {pages} {046302} (\bibinfo
  {year} {2010})}\BibitemShut {NoStop}%
\bibitem [{\citenamefont {Denniston}\ \emph {et~al.}(2004)\citenamefont
  {Denniston}, \citenamefont {Marenduzzo}, \citenamefont {Orlandini},\ and\
  \citenamefont {Yeomans}}]{marenduzzo3}%
  \BibitemOpen
  \bibfield  {author} {\bibinfo {author} {\bibfnamefont {C.}~\bibnamefont
  {Denniston}}, \bibinfo {author} {\bibfnamefont {D.}~\bibnamefont
  {Marenduzzo}}, \bibinfo {author} {\bibfnamefont {E.}~\bibnamefont
  {Orlandini}}, \ and\ \bibinfo {author} {\bibfnamefont {J.~M.}\ \bibnamefont
  {Yeomans}},\ }\href@noop {} {\bibfield  {journal} {\bibinfo  {journal} {Phil.
  Trans. Roy. Soc. Ser. A}\ }\textbf {\bibinfo {volume} {362}},\ \bibinfo
  {pages} {1745} (\bibinfo {year} {2004})}\BibitemShut {NoStop}%
\bibitem [{\citenamefont {Wood}\ \emph {et~al.}(2011)\citenamefont {Wood},
  \citenamefont {Lintuvuori}, \citenamefont {Schofield}, \citenamefont
  {Marenduzzo},\ and\ \citenamefont {Poon}}]{marenduzzo4}%
  \BibitemOpen
  \bibfield  {author} {\bibinfo {author} {\bibfnamefont {T.~A.}\ \bibnamefont
  {Wood}}, \bibinfo {author} {\bibfnamefont {J.~S.}\ \bibnamefont
  {Lintuvuori}}, \bibinfo {author} {\bibfnamefont {A.~B.}\ \bibnamefont
  {Schofield}}, \bibinfo {author} {\bibfnamefont {D.}~\bibnamefont
  {Marenduzzo}}, \ and\ \bibinfo {author} {\bibfnamefont {W.~C.~K.}\
  \bibnamefont {Poon}},\ }\href@noop {} {\bibfield  {journal} {\bibinfo
  {journal} {Science}\ }\textbf {\bibinfo {volume} {334}},\ \bibinfo {pages}
  {79} (\bibinfo {year} {2011})}\BibitemShut {NoStop}%
\bibitem [{\citenamefont {Tiribocchi}\ \emph {et~al.}(2014)\citenamefont
  {Tiribocchi}, \citenamefont {Henrich}, \citenamefont {Lintuvuori},\ and\
  \citenamefont {Marenduzzo}}]{tiribocchi3}%
  \BibitemOpen
  \bibfield  {author} {\bibinfo {author} {\bibfnamefont {A.}~\bibnamefont
  {Tiribocchi}}, \bibinfo {author} {\bibfnamefont {O.}~\bibnamefont {Henrich}},
  \bibinfo {author} {\bibfnamefont {J.~S.}\ \bibnamefont {Lintuvuori}}, \ and\
  \bibinfo {author} {\bibfnamefont {D.}~\bibnamefont {Marenduzzo}},\
  }\href@noop {} {\bibfield  {journal} {\bibinfo  {journal} {Soft Matter}\
  }\textbf {\bibinfo {volume} {10}},\ \bibinfo {pages} {4580} (\bibinfo {year}
  {2014})}\BibitemShut {NoStop}%
\bibitem [{\citenamefont {Foffano}\ \emph {et~al.}(2014)\citenamefont
  {Foffano}, \citenamefont {Lintuvuori}, \citenamefont {Tiribocchi},\ and\
  \citenamefont {Marenduzzo}}]{foffano}%
  \BibitemOpen
  \bibfield  {author} {\bibinfo {author} {\bibfnamefont {G.}~\bibnamefont
  {Foffano}}, \bibinfo {author} {\bibfnamefont {J.~S.}\ \bibnamefont
  {Lintuvuori}}, \bibinfo {author} {\bibfnamefont {A.}~\bibnamefont
  {Tiribocchi}}, \ and\ \bibinfo {author} {\bibfnamefont {D.}~\bibnamefont
  {Marenduzzo}},\ }\href@noop {} {\bibfield  {journal} {\bibinfo  {journal}
  {Liquid Crystal Reviews}\ }\textbf {\bibinfo {volume} {2}},\ \bibinfo {pages}
  {1} (\bibinfo {year} {2014})}\BibitemShut {NoStop}%
\bibitem [{\citenamefont {Tiribocchi}\ \emph {et~al.}(2016)\citenamefont
  {Tiribocchi}, \citenamefont {Da~Re}, \citenamefont {Marenduzzo},\ and\
  \citenamefont {Orlandini}}]{tiribocchi5}%
  \BibitemOpen
  \bibfield  {author} {\bibinfo {author} {\bibfnamefont {A.}~\bibnamefont
  {Tiribocchi}}, \bibinfo {author} {\bibfnamefont {M.}~\bibnamefont {Da~Re}},
  \bibinfo {author} {\bibfnamefont {D.}~\bibnamefont {Marenduzzo}}, \ and\
  \bibinfo {author} {\bibfnamefont {E.}~\bibnamefont {Orlandini}},\ }\href@noop
  {} {\bibfield  {journal} {\bibinfo  {journal} {Soft Matter}\ }\textbf
  {\bibinfo {volume} {12}},\ \bibinfo {pages} {8195} (\bibinfo {year}
  {2016})}\BibitemShut {NoStop}%
\bibitem [{\citenamefont {Cates}\ \emph {et~al.}(2009)\citenamefont {Cates},
  \citenamefont {Henrich}, \citenamefont {Marenduzzo},\ and\ \citenamefont
  {Stratford}}]{cates2}%
  \BibitemOpen
  \bibfield  {author} {\bibinfo {author} {\bibfnamefont {M.~E.}\ \bibnamefont
  {Cates}}, \bibinfo {author} {\bibfnamefont {O.}~\bibnamefont {Henrich}},
  \bibinfo {author} {\bibfnamefont {D.}~\bibnamefont {Marenduzzo}}, \ and\
  \bibinfo {author} {\bibfnamefont {K.}~\bibnamefont {Stratford}},\ }\href@noop
  {} {\bibfield  {journal} {\bibinfo  {journal} {Soft Matter}\ }\textbf
  {\bibinfo {volume} {5}},\ \bibinfo {pages} {3791} (\bibinfo {year}
  {2009})}\BibitemShut {NoStop}%
\bibitem [{\citenamefont {Carenza}\ \emph {et~al.}(2019)\citenamefont
  {Carenza}, \citenamefont {Gonnella}, \citenamefont {Lamura}, \citenamefont
  {Negro},\ and\ \citenamefont {Tiribocchi}}]{tiribocchi4}%
  \BibitemOpen
  \bibfield  {author} {\bibinfo {author} {\bibfnamefont {L.~N.}\ \bibnamefont
  {Carenza}}, \bibinfo {author} {\bibfnamefont {G.}~\bibnamefont {Gonnella}},
  \bibinfo {author} {\bibfnamefont {A.}~\bibnamefont {Lamura}}, \bibinfo
  {author} {\bibfnamefont {G.}~\bibnamefont {Negro}}, \ and\ \bibinfo {author}
  {\bibfnamefont {A.}~\bibnamefont {Tiribocchi}},\ }\href@noop {} {\bibfield
  {journal} {\bibinfo  {journal} {Eur. Phys. Jour. E}\ }\textbf {\bibinfo
  {volume} {42}},\ \bibinfo {pages} {81} (\bibinfo {year} {2019})}\BibitemShut
  {NoStop}%
\bibitem [{\citenamefont {Bentley}\ and\ \citenamefont {Leal}(1986)}]{bentley}%
  \BibitemOpen
  \bibfield  {author} {\bibinfo {author} {\bibfnamefont {B.~J.}\ \bibnamefont
  {Bentley}}\ and\ \bibinfo {author} {\bibfnamefont {L.~G.}\ \bibnamefont
  {Leal}},\ }\href@noop {} {\bibfield  {journal} {\bibinfo  {journal} {J.
  Fluid. Mech.}\ }\textbf {\bibinfo {volume} {167}},\ \bibinfo {pages} {241}
  (\bibinfo {year} {1986})}\BibitemShut {NoStop}%
\bibitem [{\citenamefont {Zaleski}\ \emph {et~al.}(1995)\citenamefont
  {Zaleski}, \citenamefont {Li},\ and\ \citenamefont {Succi}}]{zaleski}%
  \BibitemOpen
  \bibfield  {author} {\bibinfo {author} {\bibfnamefont {S.}~\bibnamefont
  {Zaleski}}, \bibinfo {author} {\bibfnamefont {J.}~\bibnamefont {Li}}, \ and\
  \bibinfo {author} {\bibfnamefont {S.}~\bibnamefont {Succi}},\ }\href@noop {}
  {\bibfield  {journal} {\bibinfo  {journal} {Phys. Rev. Lett.}\ }\textbf
  {\bibinfo {volume} {75}},\ \bibinfo {pages} {2} (\bibinfo {year}
  {1995})}\BibitemShut {NoStop}%
\bibitem [{\citenamefont {Rallison}(1984)}]{rallison}%
  \BibitemOpen
  \bibfield  {author} {\bibinfo {author} {\bibfnamefont {J.~M.}\ \bibnamefont
  {Rallison}},\ }\href@noop {} {\bibfield  {journal} {\bibinfo  {journal} {Ann.
  Rev. Fluid Mech.}\ }\textbf {\bibinfo {volume} {16}},\ \bibinfo {pages} {45}
  (\bibinfo {year} {1984})}\BibitemShut {NoStop}%
\bibitem [{\citenamefont {Hakimi}\ and\ \citenamefont
  {Schowalter}(1980)}]{hakimi}%
  \BibitemOpen
  \bibfield  {author} {\bibinfo {author} {\bibfnamefont {F.~S.}\ \bibnamefont
  {Hakimi}}\ and\ \bibinfo {author} {\bibfnamefont {W.~R.}\ \bibnamefont
  {Schowalter}},\ }\href@noop {} {\bibfield  {journal} {\bibinfo  {journal} {J.
  Fluid Mech.}\ }\textbf {\bibinfo {volume} {98}},\ \bibinfo {pages} {635}
  (\bibinfo {year} {1980})}\BibitemShut {NoStop}%
\end{thebibliography}%




\end{document}